\documentclass[12pt]{article}

\usepackage{graphicx,psfrag,epsf}
\usepackage{enumerate}
\usepackage[sort,numbers]{natbib}
\usepackage{mathtools}
\usepackage{booktabs}
\usepackage{color}
\usepackage[english]{babel} % English language/hyphenation
\usepackage[protrusion=true,expansion=true]{microtype} % Better typography
\usepackage{amsmath,amsfonts,amsthm}
\usepackage{dsfont}
\usepackage{amssymb}
\usepackage{chngcntr}
\usepackage[toc,page]{appendix}
\usepackage{textcomp}
\usepackage{url}
\usepackage{hyperref}
\usepackage{placeins}
\newcommand{\mycomment}[1]{}

\usepackage{array}
\usepackage{booktabs} % Horizontal rules in tables
\usepackage{setspace}
\usepackage{amsmath}
\usepackage{xcolor}
\usepackage{soul}
\usepackage{multirow}
\usepackage{enumitem}
\usepackage{microtype} % Slightly tweak font spacing for aesthetics
\usepackage[font = small,labelfont=bf,textfont=it]{subcaption}
\usepackage{xcolor}
\usepackage{tabularx}
\usepackage[font = small,labelfont=bf,textfont=it]{caption} % Custom captions under/above floats in tables or figures
\usepackage{footnote}
\usepackage{algorithm}% http://ctan.org/pkg/algorithms
\usepackage[noend]{algpseudocode}% http://ctan.org/pkg/algorithmicx
\usepackage{etoolbox}
\usepackage{tabularx}
\usepackage{authblk}
\usepackage{caption}
\usepackage{indentfirst}
\usepackage{enumitem}
\newlist{steps}{enumerate}{1}
\setlist[steps, 1]{label = Step \arabic*:}

\algblock{ParFor}{EndParFor}
\algnewcommand\algorithmicparfor{\textbf{for}}
\algnewcommand\algorithmicpardo{\textbf{do\ parallel}}
\algnewcommand\algorithmicendparfor{\textbf{end\ parallel\ for}}
\algrenewtext{ParFor}[1]{\algorithmicparfor\ #1\ \algorithmicpardo}
\algrenewtext{EndParFor}{\algorithmicendparfor}

\makeatletter
\g@addto@macro\normalsize{%
  \setlength\abovedisplayskip{5pt}
  \setlength\belowdisplayskip{5pt}
  \setlength\abovedisplayshortskip{5pt}
  \setlength\belowdisplayshortskip{5pt}
}
\def\BState{\State\hskip-\ALG@thistlm}

\newcommand{\distas}[1]{\mathbin{\overset{#1}{\kern\z@\sim}}}%
\newcommand{\bm}[1]{\mathbf{#1}}
\newcommand{\w}[1]{\mathcal{#1}}
\newcommand{\bs}[1]{\boldsymbol{#1}}
\newsavebox{\mybox}\newsavebox{\mysim}
\newcommand{\distras}[1]{%
  \savebox{\mybox}{\hbox{\kern3pt$\scriptstyle#1$\kern3pt}}%
  \savebox{\mysim}{\hbox{$\sim$}}%
  \mathbin{\overset{#1}{\kern\z@\resizebox{\wd\mybox}{\ht\mysim}{$\sim$}}}%
}
\newtheorem{theorem}{Theorem}

\newtheorem{prop}[theorem]{Proposition}

\newcommand{\be}{\begin{equation}}
\newcommand{\ee}{\end{equation}}
    \newcommand{\bi}{\begin{itemize}}
\newcommand{\ei}{\end{itemize}}
\newcommand{\ben}{\begin{enumerate}}
\newcommand{\een}{\end{enumerate}}

\newcommand{\cmtS}[1]{{\color{blue}{(Simon: #1)}}}

\newcolumntype{K}[1]{\geq {\centering\arraybackslash}p{#1}}
\allowdisplaybreaks

\makeatother

\let\oldbibliography\thebibliography
\renewcommand{\thebibliography}[1]{\oldbibliography{#1}
\setlength{\itemsep}{0pt}} %Reducing spacing in the bibliography.

\newcommand{\blind}{1}

\patchcmd{\footnotemark}{\stepcounter{footnote}}{\refstepcounter{footnote}}{}{}
\setcitestyle{square}

\usepackage{titlesec}
\titlespacing*{\section}{0pt}{0.2\baselineskip}{0.3\baselineskip}
\titlespacing*{\subsection}{0pt}{0.2\baselineskip}{0.3\baselineskip}

\begin{document}

\def\spacingset#1{\renewcommand{\baselinestretch}%
{#1}\small\normalsize} \spacingset{1}

\if1\blind
{
  \title{\bf ProSpar-GP: scalable Gaussian process modeling with massive non-stationary datasets}
  \small
   \author{Kevin Li\footnote{Department of Statistical Science, Duke University},\; Simon Mak$^*$}
  \maketitle
} \fi

% \author{Jean-Fran\c{c}ois Paquet}
% \affiliation{Department of Physics and Astronomy, Vanderbilt University, Nashville TN 37235.}
% \affiliation{Department of Mathematics, Vanderbilt University, Nashville TN 37235.}
% \affiliation{Department of Physics, Duke University, Durham NC 27708.}
% \author{Steffen A. Bass}
% \affiliation{Department of Physics, Duke University, Durham NC 27708.}

\if0\blind
{
  \bigskip
  \bigskip
  \bigskip
  \begin{center}
    {\LARGE\bf ProSpar-GP: scalable Gaussian process modeling with massive non-stationary datasets}
\end{center}

  \medskip
} \fi

\begin{abstract}

Gaussian processes (GPs) are a popular class of Bayesian nonparametric models, but its training can be computationally burdensome for massive training datasets. While there has been notable work on scaling up these models for big data, existing methods typically rely on a stationary GP assumption for approximation, and can thus perform poorly when the underlying response surface is non-stationary, i.e., it has some regions of rapid change and other regions with little change. Such non-stationarity is, however, ubiquitous in real-world problems, including our motivating application for surrogate modeling of computer experiments. We thus propose a new Product of Sparse GP (ProSpar-GP) method for scalable GP modeling with massive non-stationary data. The ProSpar-GP makes use of a carefully-constructed product-of-experts formulation of sparse GP experts, where different experts are placed within local regions of non-stationarity. These GP experts are fit via a novel variational inference approach, which capitalizes on mini-batching and GPU acceleration for efficient optimization of inducing points and length-scale parameters for each expert. We further show that the ProSpar-GP is Kolmogorov-consistent, in that its generative distribution defines a valid stochastic process over the prediction space; such a property provides essential stability for variational inference, particularly in the presence of non-stationarity. We then demonstrate the improved performance of the ProSpar-GP over the state-of-the-art, in a suite of numerical experiments and an application for surrogate modeling of a satellite drag simulator.

% Approximate Gaussian Process (GP) methods have emerged as a workhorse for large-scale surrogates modeling and uncertainty quantification tasks. Unfortunately, the most popular scalable approximations methods can perform poorly when the modeled response surface exhibits non-stationary features in even moderate dimensions. Current methods for scalable non-stationary regression often rely on local approximations that may generalize poorly or employ complex global models that incur high computational cost. This paper proposes a non-stationary GP regression model that aims to solve both these drawbacks. We incorporate multiple Sparse Gaussian processes models into a valid global stochastic process via Product of Experts (POE) structure. The model permits training via minibatched variational inference and capitalizes on GPU acceleration. We show that the model outperforms current state of the art GP emulation methods on simulated examples and a satellite drag surrogate modeling problem. 

\end{abstract}

\vfill

\newpage
\spacingset{1.45} % DON'T change the spacing!

\section{Introduction}

Gaussian processes (GPs; \cite{gramacy_surrogates_2020}) are a popular class of Bayesian nonparametric models. GPs have been widely applied in broad scientific and engineering applications, including rocket design \citep{yeh2018common}, cosmology \citep{kaufman2011efficient}, and high-energy physics \citep{everett2021phenomenological,ji2021graphical,ji2022multi}, primarily due to its model flexibility and closed-form predictive equations. However, one well-known limitation of GPs is that it scales poorly for massive datasets; the computation of its posterior predictive distribution requires $\mathcal{O}(n^3)$ work and $\mathcal{O}(n^2)$ memory, where $n \gg 1$ is the sample size of the large training dataset. Such bottlenecks arise due to the need for storing and inverting an $n \times n$ covariance matrix. Without appropriate modifications, this computational burden restricts $n$ to only thousands of sample points for GP training, which presents a critical limitation in the modern era of big data.

There has thus been a notable body of work aimed at tackling such computational bottlenecks for GPs. A popular class of methods in spatial statistics involves the use of covariance tapering for sparsifying the underlying covariance matrix \citep{kaufman2008covariance,kaufman2011efficient}. Another class of methods employs low rank approximations of this covariance matrix via sampling or approximation from its spectral density \citep{rahimi2007random,li2023trigonometric}. Divide-and-conquer methods have also garnered much attention; this includes \cite{park2018patchwork}, which ``stitches'' together distinct GPs on local regions using soft continuity conditions, and \cite{sun}, which leverages a multi-level modeling approach for jointly capturing global and local trends. Recent works \citep{scaledvecchia,sauer2022vecchiaapproximated} have investigated the use of Vecchia approximations for constructing a sparse {precision} matrix via nearest neighbor conditioning sets; such methods appear to yield state-of-the-art performance in applications. Of interest to us later is the class of sparse GP (or inducing point) approaches \citep{tt_inducing, fitc}, also known as Gaussian predictive process models \citep{banerjee2008gaussian} in spatial statistics. Such methods parameterize the GP using $m \ll n$ latent ``pseudo-observations'' (to be introduced later), which induce a low-rank covariance structure that can greatly reduce training and prediction costs. Inducing point approaches also allow for easy integration of stochastic variational inference and mini-batching \cite{hensman2013gaussian} for further computational efficiency, and have thus been widely used in the machine learning (ML) literature \citep{dkl, variational_inverse, sparse_opt, liu}.

% \cmtS{i've integrated most of this paragraph above. if this looks ok then comment out this paragraph.} Significant progress has been done in scaling GPs to larger data-sets. Some of the most popular methods have involved sparse covariance approximations using inducing points \citep{tt_inducing, fitc, swsp}, random features expansions \citep{sparse_spectrum, rff_pota}, structured kernel approximations \citep{kiss, sparse_grid_kiss}, and local experts \citep{bcm, cao_fleet, cpoe, distributed, trapp_dsm, zhang_importance, park2018patchwork}. While Sparse GPs using inducing points have historically been the most popular scalable GP approach, they have been known to produce low fidelity, blurry predictions. Most recently, Vecchia approximations that form a sparse approximation to the GP precision matrix have been shown to dominate other methods in terms of predictions quality and computational cost, especially in low dimensions \cite{sauer2022vecchiaapproximated, scaledvecchia, correlation_vecchia, wu_vnn}.

A notable limitation of the above approaches, however, is that the quality of its approximation can depend greatly on the presumed model of the underlying GP, which is typically taken as stationary. In applications, however, the observed data can often be highly \textit{non-stationary}: there may be regions of the prediction space where the response surface changes rapidly, and other regions where the surface is relatively smooth. This arises naturally in our motivating surrogate modeling application for satellite drag \citep{sauer2022vecchiaapproximated}, where the goal is to train an efficient predictive model for ``emulating'' expensive computer simulations of satellite bodies moving through atmospheric gases. From prior knowledge, the response surface for satellite drag is known to be highly non-stationary \citep{mehta2014modeling}, as different regions of the parameter space correspond to distinct physical regimes. With such non-stationarity in the massive training data, the aforementioned existing methods (whose approximation can greatly depend on its underlying stationary assumption) may yield a mediocre fit of the response surface, as we shall see later.

One seemingly straight-forward solution is to simply fit a GP model with a non-stationary covariance kernel; such models have been widely explored in the context of spatial statistics \citep{paciorek_nonstationary, remes_spectral, pmlr-v51-heinonen16}. However, there are difficulties in leveraging existing scalable approximation techniques for these non-stationary models, particularly in higher-dimensional settings. An alternate approach is deep GP (DGP) modeling \citep{daminou_dgp, dunlop2018deep}, for which numerous approaches have been proposed for scalable prediction, including expectation propagation \citep{dgp_ep}, doubly stochastic variational inference \citep{dsvi}, and Hamiltonian Monte Carlo \citep{dgp_hmc}. In particular, a recent work \citep{sauer2022vecchiaapproximated} utilized Vecchia approximations with elliptical slice sampling for DGP fits. Despite such techniques, DGPs can still be computationally expensive to fit for large datasets, and can have highly complex posterior geometries that make it difficult to fully explore for stable inference and prediction \citep{pleiss2021limitations}; we shall see this in later experiments.
% to DGPs in order to achieve the high fidelity predictions required for the low noise to signal ratio setting prevalent in computer experiments
% ifficult to perform inference/diagnostics on, and degenerate to shallow GPs with moderate layer widths 

% In this paper we provide an novel alternative strategy for implementing non-stationary GP regression in higher dimensions that

% retains the flexibility and scalability of DGPs without the many theoretical and computational pathologies

 % We propose incorporating $J$ sparse GP processes with distinct kernel hyper-parameters and inducing points into a single stochastic process using a Product of Expert (POE) approach.

We thus propose a new Product of Sparse GP (ProSpar-GP) method that addresses the above challenges for scalable GP training with massive non-stationary datasets. The key idea is to leverage a ``product-of-experts'' formulation \citep{hinton2002training} of sparse GP experts, which aggregates $J \geq 1$ sparse GP experts into a global probabilistic predictive model. This is achieved by placing different experts (with distinct kernel hyperparameters and inducing points) within local regions to account for non-stationarity over the prediction space. With a carefully-designed variational inference procedure, each sparse GP expert can be trained via an efficient optimization of its length-scale parameters and local inducing points. Our variational inference procedure can be performed in $\mathcal{O}(\sum_{j=1}^J m_j^3)$ runtime and $\mathcal{O}(\sum_{j=1}^J m_j^2)$ memory, where $m_j \ll n$ is the number of inducing points for the $j$-th sparse GP expert. The developed procedure can further capitalize on mini-batching and GPU acceleration for computational efficiency. With this, we then demonstrate the improved predictive performance of the ProSpar-GP over the state-of-the-art, in a suite of numerical experiments and the motivating surrogate modeling application.

% We further show that the generative model underlying the ProSpar-GP corresponds to a valid stochastic process (Kolmogorov consistency \citep{}), which enables principled regularization and hyper-parameter learning via a global marginal likelihood.

It is worth noting that, in the ML literature, there is a body of work on product-of-expert (POE) formulations with \textit{standard} GP experts. This was first explored in \cite{cao_fleet}, then extended in \cite{distributed} for efficient computation. One known limitation is that such approaches typically yield poorly calibrated uncertainties, due to its reliance on an invalid joint probability distribution. \cite{rbcm, cohen_heal} explored alternate expert aggregation strategies to address this; however, the resulting generative model of such methods does not define a valid stochastic process \citep{samo2016string}, which (as we shall see later) can yield poor predictive performance and uncertainty quantification. The proposed ProSpar-GP has a key advantage over the above methods: we prove that our approach is Kolmogorov-consistent, in that its generative distribution defines a valid stochastic process over the prediction space. This permits stable variational inference and prediction under our model \citep{lotfi2022bayesian}, and allows for greater modeling flexibility for each GP expert to better capture non-stationary behavior (more on this in Section \ref{sec:recent}). Due in part to this stability, we show later in a suite of numerical experiments that the ProSpar-GP offers improved performance over existing POE approaches. 

% and the authors restricted themselves to experts with homogeneous length-scales to prevent over-fitting \citep{distributed, cohen_heal}

The paper is organized as follows. Section \ref{sec:background} provides a brief review of GPs, sparse GPs and other state-of-the-art methods, and explores their potential limitations for massive non-stationary datasets. Section \ref{sec:prospar} presents the proposed ProSpar-GP, including the employed variational inference approach that leverages mini-batching and GPU acceleration for scalable prediction. Section \ref{sec:prop} investigates important properties of the ProSpar-GP, including computational and memory complexity, Kolmogorov consistency, and global-local modeling. Section \ref{sec:exp} compares the ProSpar-GP to the state-of-the-art in a suite of numerical experiments. Section \ref{sec:app} explores its performance in an application to surrogate modeling of a satellite drag simulator. Section \ref{sec:conc} concludes with final thoughts.

% This Product of Sparse Gaussian Process Experts (ProSpar) model permits more efficient use of inducing points as low expert length-scales in a local region will not require a higher density of inducing points globally. points.  Unlike current local experts models our approach 1) trains each expert on the entire input space 2)  admits Kolmogorov consistent stochastic process. We develop a stochastic variational inference scheme that achieves $\w{O}(\sum_{j=1}^J m_j^3)$ time and $\w{O}(\sum_{j=1}  m_j^2)$ per iteration memory complexity. In addition, a significant portion of computational burden involves Matrix Vector Multiplications which allows extremely efficient use of GPU acceleration. Using numerical experiments and a real-world emulation problem, we show that ProSpar-GP is faster than the state of the art while providing superior probalistic predictions. 

%Each sparse GP expert $j$ is equipped with distinct kernel hyper-parameters and has $m_j \ll \sum_{j=1}^{J}$ inducing points initialized in local regions of data.  In addition to more efficient use of inducing points, this structure allows the use of of Stochastic variational inference to achieve  $\w{O}(\sum_{j=1}^J m_j^3)$ time and $\w{O}(\sum_{j=1}  m_j^2)$ memory complexity. We will show through simulated examples and a real-world emulation task that ProSpar-GP outperforms state of the art GP-regression methods on non-stationary regression tasks in higher dimensions. 

\section{Background \& Motivation}
\label{sec:background}

\subsection{Gaussian process modeling}

Gaussian processes \citep{gpml, gramacy_surrogates_2020} are a widely-used class of Bayesian non-parametric models. Suppose we wish to predict an unknown scalar function $f(\mathbf{x})$ over an input domain $\mathcal{X} \subseteq \mathbb{R}^d$. A (zero-mean) GP prior on $f(\cdot)$ takes the form $f(\cdot) \sim \text{GP}\{0,k_{\boldsymbol{\phi}}(\cdot,\cdot)\}$, where $k_{\boldsymbol{\phi}}(\cdot,\cdot)$ is a pre-specified covariance function with length-scale parameters $\boldsymbol{\phi}$. Suppose we observe data $\mathbf{y} = (y_1, \cdots, y_n)$, following:
\begin{equation}
y_i = f(\mathbf{x}_i) + \epsilon_i, \quad i = 1, \cdots, n,
\end{equation}
where $\epsilon_i \distas{i.i.d.} \mathcal{N}(0,\gamma^2)$ are i.i.d. normal noise terms. Conditioning on such data, one can show \citep{gpml} that the posterior distribution of $f(\mathbf{x}_{\rm new})$ at a new input $\mathbf{x}_{\rm new}$ takes the form:
\begin{equation}
f(\bm{x}_{\rm new}) | y_1, \cdots, y_n \sim \w{N}\{\mu(\mathbf{x}_{\rm new}), \sigma^2(\mathbf{x}_{\rm new})\},
\label{eq:gppred}
\end{equation}
where:
\small
\begin{equation}
\mu(\mathbf{x}_{\rm new})= \mathbf{k}^T_n(\bm{x}_{\rm new}) (\bm{K}_{n,n} + \gamma^2\bm{I})^{-1}\bm{y}, \; \sigma^2(\mathbf{x}_{\rm new})= k_{\phi}(\mathbf{x}_{\rm new},\mathbf{x}_{\rm new}) - \mathbf{k}^T_n(\bm{x}_{\rm new})  (\bm{K}_{n,n} + \gamma^2\bm{I})^{-1} \mathbf{k}_n(\bm{x}_{\rm new})
\label{eq:gppredeqn}
\normalsize
\end{equation}
\noindent are its posterior mean and variance expressions. Here, $\bm{K}_{n,n} = [k_\phi(\bm{x}_i,\bm{x}_j)]_{i,j=1}^n$ is the $n \times n$ covariance matrix for the data, and $\bm{k}_n(\bm{x}_{\rm new}) = (k_\phi(\bm{x}_i,\bm{x}_{\rm new}))_{i=1}^n$ is the covariance vector between the data and the new input $\bm{x}_{\rm new}$. Equation \eqref{eq:gppred} thus provides a closed-form expression for the GP posterior predictive distribution. There are various ways for fitting the required model parameters $\boldsymbol{\phi}$ and $\gamma^2$, including maximum likelihood estimation, empirical Bayes and fully Bayesian inference; see \cite{gramacy_surrogates_2020} for details.

Equation \eqref{eq:gppredeqn} reveals the primary computational bottleneck for GP modeling with big data: the storage and inverse computation of the $n \times n$ matrix $\bm{K}_{n,n} + \gamma^2\bm{I}$ require $\mathcal{O}(n^2)$ memory and $\mathcal{O}(n^3)$ work, respectively. This can clearly be computationally infeasible when the sample size $n$ is larger than several thousands. This issue is further compounded by the need to estimate the model hyperparameters $\boldsymbol{\phi}$ and $\gamma^2$ from data. Using either a maximum likelihood or a Bayesian approach, this estimation typically requires many evaluations of its likelihood function, with each evaluation involving a separate inverse of the $n \times n$ covariance matrix. Without modifications, this thus greatly limits the use of GP models for massive training datasets.

% If a latent function $f(\bm{x})$ is drawn from a GP prior, any finite realization $\bm{f} = \{f_i\}_{i=1}^n$ observed at points $\bm{X} = \{\bm{x_i}\}_{i=1}^n$ will be distributed $\bm{f} \sim \w{N}(0, \bm{K}_{nn})$ where $\bm{K}_{nn}^{i,j} = k(\bm{x}_i, \bm{x}_j)$ and $k_{\phi}(\cdot, \cdot)$ is a kernel function equipped with hyper-paramters $\phi$.In the regression set up where our observations follow a gaussian likelihood $y_i = f(\bm{x}_i) + \epsilon_i \ , \epsilon_i \sim N(0, \sigma^2)$,

% we can optimzie hyperparamters $\phi, \sigma^2$ by maximizing the marginal likelihood
% $$
% \log\left( p(\bm{y} |\bm{X}) \right) = \log\left( \w{N}(\bm{y}| 0, \bm{K}_{nn} + \sigma^2 \bm{I}) \right) 
% $$
% Once the hyperparameters have been optimized, we can calculated predictive distribution on unseen points $\bm{X}^{*}$ induced by our GP prior:

% Computation of the marginal likelihood requires $\w{O}(n^3)$ operations due to inversion of the GP covariance matrix $(\bm{K}_{nn} + \sigma^2I)^{-1}$. Even prediction requires $\w{O}(n^2)$ due to the matrix multiplication $K_{\bm{X}^{*}, \bm{X}^{*}} - \bm{K}_{\bm{X}^{*}, \bm{X}} (\bm{K}_{nn} + \sigma^2I)^{-1}$. Both these operations become intractable for data sets larger than a few thousand observations. 

\subsection{Sparse Gaussian processes}
\label{sec:sgp}

A popular strategy on tackling this limitation in the ML literature is via sparse Gaussian processes \citep{hensman2013gaussian, fitc, tt_inducing}; this is also known as Gaussian predictive processes in spatial statistics \citep{banerjee2008gaussian}. The key idea is to leverage a set of $m \ll n$ ``representative'' \textit{inducing points} $\{\mathbf{z}_l\}_{l=1}^m \subseteq \mathcal{X}$ over the input space, where the number of inducing points $m$ is much smaller than the sample size $n$. Let $\mathbf{f} = (f(\mathbf{x}_i))_{i=1}^n$ be the latent function values at training input points, and $\mathbf{u} = (f(\mathbf{z}_l))_{l=1}^m$ be the latent \textit{pseudo-observations} at inducing points. Following the fully-independent-training-conditional (FITC) formulation in \cite{fitc}, we adopt the approximation that the entries in $\mathbf{f}$ are conditionally independent given the pseudo-observations $\mathbf{u}$. This can be represented hierarchically as:
\begin{align}
\begin{split}
\bm{y}|\mathbf{f} &= \bm{f} + \bs{\epsilon} , \quad \bs{\epsilon} \sim \w{N}(0, \gamma^2\bm{I}), \\
\bm{f}|\mathbf{u} &\distas{i.i.d.} \w{N}( \bm{K}_{n,m} \bm{K}_{m,m}^{-1}  \bm{u}, \bs{\Lambda}), \quad \bs{\Lambda} = \mbox{diag}\{ \bm{K}_{n,n} - \bm{K}_{n,m}^T\bm{K}_{m,m}^{-1}\bm{K}_{m,n}\},\\
\bm{u} &\sim  \w{N}(0, \bm{K}_{m,m}).
\label{eq:fitc}
\end{split}
\end{align}
Here, $\bm{K}_{m,m} = [k_\phi(\bm{z}_l,\bm{z}_{l'})]_{l,l'=1}^m$ is the $m \times m$ covariance matrix at inducing points, and $\bm{K}_{n,m} = {[k_\phi(\bm{x}_i,\bm{z}_l)]_{l=1}^m}_{i=1}^n$ is the $n \times m$ cross-covariance matrix between inducing points and observed inputs.

With this approximation, the corresponding marginal likelihood $p(\bm{y})$ and the posterior distribution of pseudo-observations $[\bm{u}| \bm{y}]$ admit efficient closed-form expressions. The kernel parameters $\boldsymbol{\phi}$, noise variance $\gamma^2$ and inducing points $\{\mathbf{z}_l\}_{l=1}^m$ can then be estimated via maximization of this closed-form marginal likelihood, for which each evaluation requires $\mathcal{O}(nm^2 + m^3)$ work. Finally, with these parameters fitted, the desired (approximate) predictive posterior distribution $[f(\bm{x}_{\rm new})|\bm{y}]$ can be computed via marginalization, i.e., $[f(\bm{x}_{\rm new})|\bm{y}] = \int [f(\bm{x}_{\rm new})|\bm{u}][\bm{u}|\bm{y}] d\bm{u}$, where $[\bm{u}|\bm{y}]$ again can be obtained in closed-form from \eqref{eq:fitc}. This marginalization yields a closed-form distribution for $[f(\bm{x}_{\rm new})|\bm{y}]$ that can be computed in $\mathcal{O}(nm^2 + m^3)$ work, which greatly reduces the $\mathcal{O}(n^3)$ work needed for standard GPs. There has since been many extensions of sparse GPs, including the use of variational inference for efficient estimation of model parameters; see \citep{tt_inducing, hensman2013gaussian}.

% Sparse GP methods rely on similar methods often substituting the marginal likelihood for the Evidence Lower Bound (ELBO). 

 % which allow for inference and prediction wit $\w{O}(nm^2 + m^3)$ operations

% have been an extremely popular and successfully method for reducing the $\w{O}(n^3)$ scaling associated with GP regression. A significant contributor of their success is that they make extremely efficient use of GPU resources , as all necessary quantities for inference and prediction can be computed using Matrix Vector Multiplication \citep{black_box, tresp}. Many different variants exists, but the essential idea is to use $m \ll n$ inducing points to parametrize a low rank approximation to the full covariance matrix $\bm{K}_{nn}$. Although the variational formulation proposed by \citep{vgp, hensman2013gaussian} has become preferred \citep{bauer, burt}, ProSpar will rely on the formulation known as the Fully Independent training conditional (FITC). FITC assumes that the latent function values $\bm{f}$ are independent conditional on the $m$ inducing points $\bm{u}$ obvservaed at locations $\bm{Z} = \{\bm{z}_{l}\}_{l=1}^m$ such that

However, a key limitation of existing sparse GP approaches is that, for a fixed number of inducing points $m$, its approximation of the desired ``full'' GP fit (i.e., using all $n$ data points) can become increasingly poor when the response surface is non-stationary; such non-stationarity, however, is ubiquitous in many applications. Non-stationarity introduces two critical challenges for existing sparse GP methods. First, since the representative inducing points $\{\mathbf{z}_l\}_{l=1}^m$ are optimized via the marginal likelihood of a presumed \textit{stationary} GP, these optimized points can thus be far away from regions of high activity in the response surface. With such points, the resulting sparse GP can then yield a poor approximation of the full GP fit. Second, it is known \citep{cole2021locally,burt} that model misspecification (in the form of non-stationarity) can cause highly unstable estimates of length-scale parameters. One solution might be to increase the number of inducing points $m$ in hopes of better capturing non-stationary regions in the data. This, however, requires a large choice of $m$, which greatly diminishes the computational efficiency of sparse GPs.

% This inducing point specification can become computationally wasteful when the underlying function violates stationarity assumptions and varies quickly only in local regions of the input space \citep{input_dependent_sparse}. Low-lengthscales allow for more flexible modeling of fast varying functions by reducing the region in which inducing points are informative. Therefore, low length-scales necessary to model local variation may require a computationally infeasible number of inducing points to maintain globally precise predictions. Furthermore, balancing between modeling areas of fast and slow variation with a inducing point budget creates pathologies when optimizing lengthscales and inducing point locations \citep{cole2021locally, burt}.

\begin{figure}[t]
\centering
\subfloat[]{%
  \includegraphics[width=5cm]{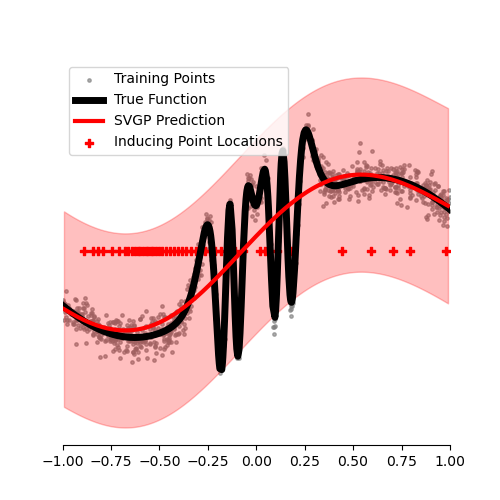}%
  \label{fig:illustrate:optimize}%
}
\subfloat[]{%
  \includegraphics[width=5cm]{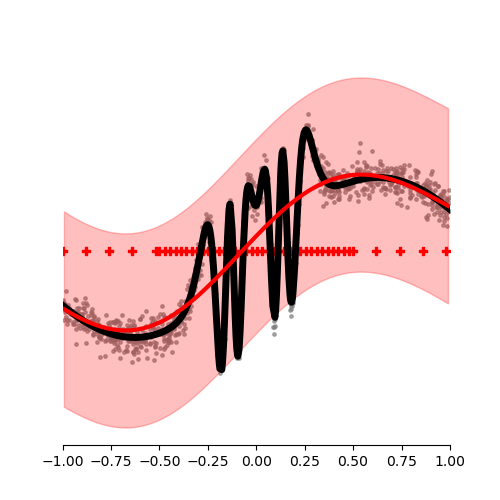}%
  \label{fig:illustrate:manual}%
} 
\subfloat[]{%
  \includegraphics[width=5cm]{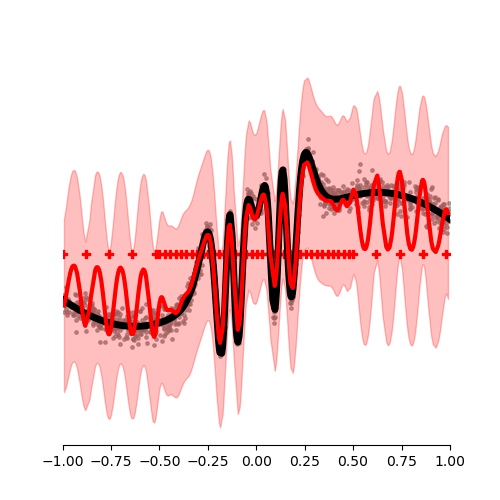}%
  \label{fig:illustrate:ls}%
}
\label{fig:ill}
\caption{Visualizing three different sparse GP fits on a 1-d example with $n=750$ observations and $m=45$ inducing points. The posterior mean predictor is shown as the solid red curve, with its 95\% predictive intervals shaded in red. (a, left) uses the length-scale parameter and inducing points optimized via the marginal likelihood. (b, middle) uses the optimized length-scale and manually-placed inducing points. (c, right) uses the same inducing points in (b) with a small choice of length-scale parameter.
% Different Sparse-GP model specifications trained on non-sttionary toy function with $n= 750$ observations and $m = 45$ inducing points. \ref{fig:illustrate:optimize} length-scales and inducing points locations optimzied via marginal likelihood; \ref{fig:illustrate:manual} optimized length-scales and inducing points manually placed in area of fast variation;  \ref{fig:illustrate:ls} manual inducing point placement and low length-scales. \ref{fig:illustrate:mine} ProSpar GP. Each color indicates that the group of inducing points belongs to a distinct expert with unique kernel hyper-parameters. Inducing point locations are optimized.
}
\label{fig:illustrate}
\end{figure}

We use a simple $d=1$ example to illustrate such limitations. Figure \ref{fig:illustrate} shows the true function $f(\cdot)$ in black, chosen as a highly non-stationary function that varies quickly within the region $[-0.5,0.5]$ and slowly otherwise. We then generate $n=750$ observations, with noise standard deviation set as $15\%$ of the standard deviation of function values. With this data, we then fit several sparse GPs with $m=45$ inducing points (marked as red points in the figure), using the squared-exponential kernel. The first, (a), makes use of \textit{optimized} length-scale parameter $\boldsymbol{\phi}$ and inducing points $\{\bm{z}_l\}_{l=1}^m$ via maximization of the marginal likelihood from a \textit{stationary} GP, as recommended in \cite{fitc, hensman2013gaussian, tt_inducing}. We see that these inducing points are not concentrated in the region of high variation, i.e., $[-0.5, 0.5]$, which results in poor predictive performance and overly conservative uncertainty quantification. The second model, (b), aims to correct this by manually placing more inducing points within $[-0.5,0.5]$, with length-scale again optimized via the marginal likelihood from a stationary GP. Despite this, we see that the estimated length-scale is much too high to adequately capture the high variation within the middle region. Finally, the third model (c) tries to address this by imposing a much smaller upper bound for length-scale optimization, using the same inducing points in (b). With this smaller length-scale (along with high density of inducing points within $[-0.5,0.5]$), we see much better predictions within the region of high variation. However, this comes at a clear cost: with a small global length-scale parameter, its prediction within regions of low variation now becomes highly volatile and inaccurate. Such issues are exacerbated in the presence of non-stationarity in \textit{higher} dimensions, where there can be many local regions of high variation that can be identified with massive training data.

\subsection{Recent developments for scalable GPs}
\label{sec:recent}

% The quality of the approximation can be very high for $m \ll n$ \cite{Katzfuss_2021}.

We now briefly review recent developments and the state-of-the-art, including its potential limitation for modeling massive non-stationary datasets. Particularly in the ML literature, there has been many extensions of sparse GPs \citep{liu, bauer}, including variational approximations \citep{tt_inducing, hensman2013gaussian}, hybrid MCMC \citep{optimize_integrate}, and amortized implementations \citep{input_dependent_sparse}. Such methods, however, have been noted to suffer from blurry (i.e., over-smoothed) predictions, due to its approximation via a stationary model using few inducing points \citep{burt, wu_vnn, sauer2022vecchiaapproximated}. A recent promising development is the class of Vecchia-approximated GPs, which approximates the underlying precision matrix by conditioning each observation on a subset of $q$ data points (typically its $q$-nearest neighbors). This approximation induces a sparse precision matrix, which allows for efficient inference of model parameters (e.g., length-scales) and subsequent prediction in $\w{O}(nq^3)$ work. One potential limitation of Vecchia methods is that they can be highly sensitive to the choice of conditioning sets and ordering. Conditioning sets are typically selected via scaled Euclidean distance metrics, which may encounter difficulties for non-stationary data in moderate dimensions, as we shall see later. \cite{correlation_vecchia} tackles this via the construction of conditioning sets and orderings with a distance function learned from a non-stationary covariance kernel. This, however, requires the learning of such a non-stationary kernel, which is again difficult in moderate dimensions and with large sample sizes \cite{sauer2023nonstationary}; such methods have thus been largely applied in lower-dimensional spatio-temporal settings. 

Deep GPs (DGPs; \citep{daminou_dgp, dunlop2018deep}) offer an alternative approach, by warping the input space through multiple GP layers to model potential non-stationarity. A popular approach for scalable DGP fitting is via Doubly Stochastic Variational Inference (DSVI) \citep{dsvi}, which extends the earlier inducing points idea; this approach suffers from similar limitations, however, in that limited inducing points restrict the expressiveness of the model \citep{sauer2022vecchiaapproximated}. A recent promising approach \citep{sauer2022vecchiaapproximated} aims to address this via the combination of Vecchia approximations with DGPs. However, while DGPs offer greater expressiveness over standard GPs, such models are inherently difficult to fit: their posteriors are inherently multi-modal, with complex symmetries that can make parameter inference and subsequent predictions highly unstable \citep{implicit_dgp, iwp}. This becomes further pronounced in the presence of massive non-stationary training data, as we show later. The theoretical underpinnings of DGPs are also scant, and recent work has shown that DGPs may easily degenerate to stationary GPs in practical problems \citep{thinanddeep, pleiss2021limitations}. 

% In addition, DGPs are known to quickly degenerate to shallow GPs as the widths of the intermediate layers increase \citep{pleiss2021limitations}. 

% In addition, it is uncertain whether their Vecchia conditioning scheme preserves full DGP representations as fixing conditioning sets during training periods may limit learning. 

% . Local expert models partition the data set between separate GP experts that can be equipped with distinct kernel hyper-parameters in order to accommodate non-stationarity

% These methods provide efficient and often easily parrlelizable computational schemes.

% Gramacy alleviates these issues by combining local and global information to train local predictive experts at each test point based on local designs. 

% extends this method by incorporating global information into the construction of the local designs

Finally, another class of methods involves local GP experts, which leverage local GP fits on partitions of the data to better model non-stationarity in a scalable manner. This includes the local GP approximations in \cite{lagp} and its extension \cite{sun}. \cite{park2018patchwork} leverages a generative model of local GP models patched together with loose continuity conditions. Such local methods, however, may overfit to local partitions and disregard global structure, which can harm generalization in moderate dimensions \citep{liu}. A recent development on this front is \cite{cohen_heal}, which uses a ``healed'' generalized product of local experts (HgPOE); each expert is a GP fitted on a disjoint partition of the big data, then aggregated by taking the normalized product of predictive densities over all $J \geq 1$ experts. This builds on a rich literature on product-of-experts modeling; see the seminal work \cite{hinton2002training} and its extensions \citep{tresp, rbcm, cohen_heal, cpoe, bcm}. More specifically, the HgPOE adopts the following approximation for the predictive density of $f(\bm{x}_{\rm new})$:
\begin{align}
&p(f(\bm{x}_{\rm new})| \bm{y}) \propto \prod_{j=1}^J \phi(f(\bm{x}_{\rm new});\mu_{j}(\bm{x}_{\rm new}), \sigma^2_{j}(\bm{x}_{\rm new}))^{\alpha_j(\bm{x}_{\rm new})}.
\label{eq:hgpoe}
\end{align}
Here, $\phi(\cdot;\mu,\sigma^2)$ is the normal density with mean $\mu$ and variance $\sigma^2$, and $\mu_{j}(\bm{x}_{\rm new})$ and $\sigma^2_{j}(\bm{x}_{\rm new})$ are the posterior mean and variance of the $j$-th GP expert (see \eqref{eq:gppredeqn}) fit using only data from the $j$-th data partition. The function $\alpha_j(\bm{x}_{\rm new})$ controls the \textit{influence} of expert $j$ in the aggregate prediction, with a larger $\alpha_j$ dictating greater influence. Softmax weights of the form $\alpha_j(\bm{x}_{\rm new}) = {\exp\{-T \sigma^2_{j}(\bm{x}_{\rm new})\}}/{\sum_{k=1}^J \exp\{-T \sigma^2_{k}(\bm{x}_{\rm new})\}}$ are recommended in \cite{cohen_heal}, where $T > 0$ is a pre-selected hyperparameter that controls the smoothness of expert weights. One key limitation of the HgPOE is that it is Kolmogorov-inconsistent \citep{samo2016string}: its aggregation does not define a valid stochastic process on $f(\cdot)$. This is likely a cause for the observed instability of such methods \citep{distributed, cao_fleet, cohen_heal, trapp_dsm}, particularly in its tendency to overfit when allowing expert-specific length-scale parameters \citep{distributed, cohen_heal}.

\subsection{Illustrating example}
\begin{figure}[!t]
\centering
\subfloat[Visualizing the 2-d Michalewicz function (image from \url{https://www.sfu.ca/~ssurjano/michal.html}).]{%
  \includegraphics[width=0.45\textwidth]{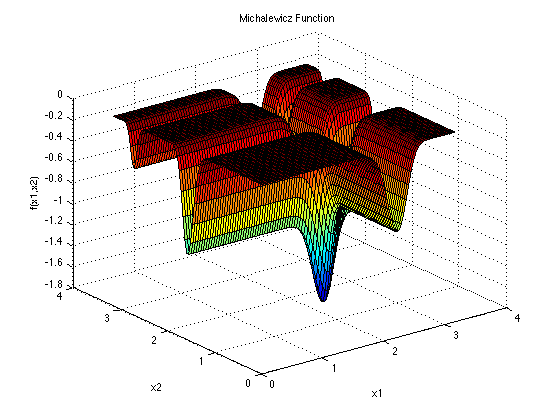}%
  \label{fig:michal:michal}%
}
\hfill
\subfloat[Test RMSEs for the compared methods using $n=100,000$ noisy samples.]{%
  \includegraphics[width=8cm]{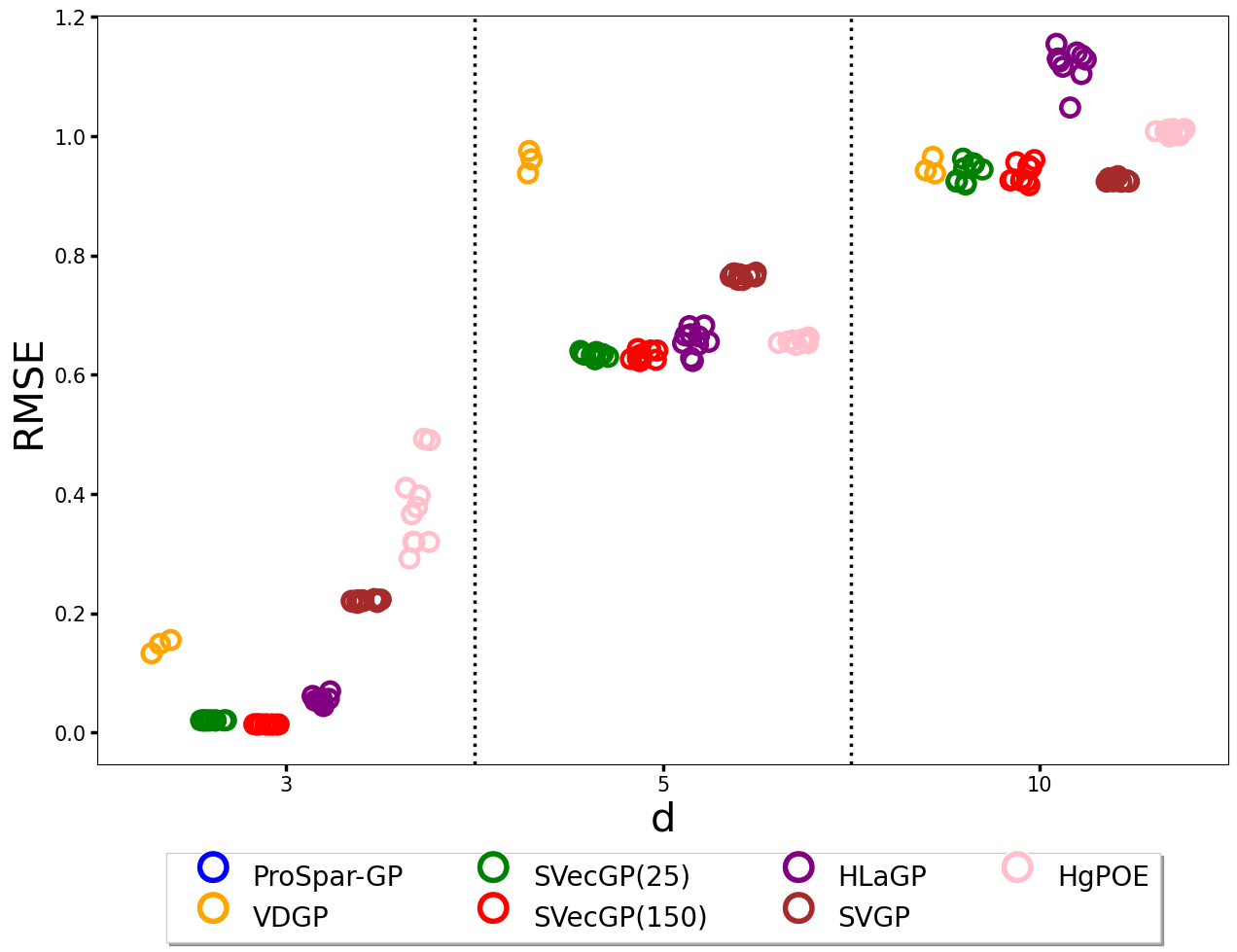}%
  \label{fig:michal:teaser}%
} 
\caption{Illustrating example with the $d$-dimensional Michalewicz function.}
\label{fig:michal}
\end{figure}

We demonstrate these potential drawbacks of the above state-of-the-art using the $d$-dimensional Michalewicz test function \citep{sun, MARREL2009742, simulationlib} (see Figure \ref{fig:michal:michal}), with inputs $\bm{x} \in [0, \pi ]^d$ in dimensions $d = 3$, $5$ and $10$. This function is highly non-stationary, with steep valleys interspersed with flat planes. We first generate a large amount ($n = 100,000$) of training data using a Latin hypercube design \citep{lhc}, with a small noise variance of $\gamma^2 = 10^{-8}$. We then compare the performance of a suite of existing methods: (i) stochastic variational inducing points (SVGP; \cite{tt_inducing}), a popular inducing points approach in ML, (ii) the state-of-the-art for Vecchia GPs: the scaled Vecchia-approximated GP (SVecGP; \citep{scaledvecchia}) and the Vecchia-approximated deep GP (VDGP; \citep{sauer2022vecchiaapproximated}), and (iii) recent local GP methods: the hybrid local approximate GP (HLaGP; \citep{sun}) and the HgPOE \citep{cohen_heal}. Predictive performance is evaluated on a separate test set of $n^*=25,000$ samples (also generated using a Latin hypercube design). All methods were compared using recommended settings using available code packages; further implementation details can be found in Section \ref{sec:comp_imp}. 

 % and $n^{*} = 25,000$ test samples from the Michaelwiz function

Figure \ref{fig:michal:teaser} shows the resulting test root-mean-squared-errors (RMSEs) of the compared methods in various dimensions $d$. While some methods generally perform well in low dimensions ($d=3$), we see that in the presence of non-stationarity, their performance can quickly deteriorate in higher dimensions ($d=5$ and $d=10$). As expected, the inducing points approach (SVGP) yields mediocre performance, as its inducing points can be far from the non-stationary regions of interest. A similar sensitivity to non-stationarity can be observed for Vecchia methods as dimension increases. While the local GP methods (HLaGP and HgPOE) aim to capture local variability via local experts, such methods appear to similarly deteriorate in performance for $d \geq 5$, which is inline with observed generalization issues in moderate dimensions \citep{liu}. This suggests that, when non-stationarity is present, the current state-of-the-art for scalable GPs may encounter difficulties in moderate to high dimensions; we aim to address this next.

% \textcolor{red}{Modified to specify only HLaGP is nonstationary, HgPOE has share lengthsales between experts}

% We also see that VDGP does not outperform the vanilla Vecchia approximation. We suspect this result is due to the ability of the vanilla Vecchia approximation to frequently update conditioning sets during length scale optimization. The frequency of VDGP condionting set updates is limited by the need to maintain good MCMC behavior (changing conditioning sets changes the model) and the inherent difficulty of learning higher dimensional warpings.

\section{The ProSpar-GP}
\label{sec:prospar}
We now introduce the proposed ProSpar-GP, which leverages the product of sparse GP experts for scalable GP training with large non-stationary data. We first introduce our model specification, then present an efficient variational inference approach for model fitting and prediction, capitalizing on mini-batching and GPU acceleration for further scalability.

\subsection{Model specification}
% now aim to construct a simple non-stationary model that is able to localize lengthscales while retaining global information/structure. 
% \textcolor{red}{In our implementation, the locality of each expert is determined by the placement of the inducing points which are initialized in local regions of the input space via an algorithm such as K-means}
To achieve effective and scalable modeling with large non-stationary data, the ProSpar-GP makes use of a carefully-constructed generative model for the observations $\bm{y}$ at observed inputs $\bm{x}_1, \cdots, \bm{x}_n$. Suppose we have $J \geq 1$ sparse GP experts, each modeling for local regions of non-stationarity over the prediction space. Here, the $j$-th expert has its distinct kernel $k^{[j]}(\cdot,\cdot)$ equipped with separate length-scale parameters $\bs{\theta}^{[j]}$, along with its own set of inducing points $\{\bm{z}_l^{[j]}\}_{l=1}^{m_j}$ and corresponding pseudo-observations $\bm{u}^{[j]} \in \mathbb{R}^{m_j}$. We will discuss how such inducing points can be optimized later in Section \ref{sec:var}. With this, the ProSpar-GP leverages the following hierarchical generative model:
\begin{align}
\bm{y}|\mathbf{f} &= \bm{f} + \bs{\epsilon} , \quad \bs{\epsilon} \sim \w{N}(0, \gamma^2\bm{I}), \quad \bm{f} = (f(\bm{x}_1), \cdots, f(\bm{x}_n)), \label{eq:spec1}\\
p(\bm{f}|\bm{u}^{[1]}, \cdots, \bm{u}^{[J]}) &\propto \prod_{j=1}^J \left\{ \prod_{i=1}^n p(f(\bm{x}_i)|\bm{u}^{[j]})^{\alpha_j(\bm{x})} \right\}, \label{eq:spec2}\\
\begin{split}
f(\bm{x}_i)|\bm{u}^{[j]} &\distas{i.i.d.} \w{N}( \bm{K}_{n,m_j}^{[j]} \bm{R}_j^{-1} \bm{u}^{[j]},  \bs{\Lambda}_{j}), \\
\bs{\Lambda}_{j} &= \text{diag}\{\lambda_j(\bm{x}_i)\}_{i=1}^n, \quad
\lambda_j(\bm{x}) = k^{[j]}(\bm{x},\bm{x}) - \bm{k}_{m_j}^{[j]}(\bm{x})^T\bm{R}_j^{-1}\bm{k}^{[j]}_{m_j}(\bm{x}),
\label{eq:spec3}
\end{split}\\
\bm{u}^{[j]} &\sim  \w{N}(\mathbf{0}, \bm{R}_j), \quad \bm{R}_j = \bm{K}_{m_j,m_j}^{[j]} + \bm{D}_j, \quad j = 1, \cdots, J. \label{eq:spec4}
% \label{eq:prospar}
\end{align}

While the above may seem quite involved, its intuition is straight-forward; we inspect below this hierarchical specification line-by-line. Equation \eqref{eq:spec1} specifies the distribution of the observed responses $\bm{y}$ given latent function values $\bm{f}$. Equation \eqref{eq:spec2} adopts a ``product-of-experts'' approximation \citep{hinton2002training,cohen_heal} of the distribution for $\bm{f}$, via the aggregation of predictive densities from the $J$ sparse GP expert. Equation \eqref{eq:spec3} then specifies the separate sparse GP models for each expert $j$, following the FITC formulation \eqref{eq:fitc}, with its corresponding inducing points $\{\bm{z}_l^{[j]}\}_{l=1}^{m_j}$ and pseudo-observations $\bm{u}^{[j]}$. Finally, each set of pseudo-observations $\bm{u}^{[j]}$ follows the marginal normal distribution from the underlying GP for expert $j$, where $\bm{K}_{m_j,m_j}^{[j]}$ is its covariance matrix and $\bm{D}_j = \text{diag}\{d_{1,j}, \cdots, d_{m_j,j}\}$ is a diagonal matrix of nugget terms \citep{peng2014choice}; we will discuss the specification of $\bm{D}_j$ later.

A key advantage of the product-of-expert aggregation in Equation \eqref{eq:spec2} is that it provides a \textit{closed-form} distribution for the latent responses $\bm{f}|\bm{u}^{[1]}, \cdots, \bm{u}^{[J]}$, which facilitates efficient variational inference and prediction. In particular, one can show that:
\begin{align}
p(\bm{f} |\bm{u}^{[1]}, \dots, \bm{u}^{[J]}) \propto \prod_{j=1}^J \left\{ \prod_{i=1}^n p(f(\bm{x}_i)|\bm{u}^{[j]})^{\alpha_j(\bm{x})} \right\} \propto \phi \left( \bm{f}; \bs{\Lambda} \sum_{j=1}^J \bm{A}_j\bs{\Lambda}_j^{-1} \bs{\mu}_j, \bs{\Lambda} \right),
\label{eq:closed}
\end{align}
where, with $\lambda(\bm{x}_i) = \left(\sum_{j=1}^J {\alpha_j(\bm{x}_i)}/{\lambda_{j}(\bm{x}_i)}\right)^{-1}$, we have $\Lambda = \text{diag}\{\lambda(\bm{x}_i)\}_{i=1}^n$, $\bs{\Lambda}_j = \text{diag}\{\lambda_{j}(\bm{x}_i)\}_{i=1}^n$, $\bm{A}_j =  \text{diag}\{\alpha_j(\bm{x}_i)\}_{i=1}^n$, and $\bs{\mu}_j = \bm{K}_{n,m_j}^{[j]} \bm{R}_j^{-1} \bm{u}^{[j]}$. In other words, provided the pseudo-observations from all $J$ experts, the conditional distribution of $\bm{f}$ after expert aggregation reduces to a (closed-form) multivariate normal distribution. Taking \eqref{eq:closed} along with the marginal distribution of pseudo-observations $\bm{u}^{[1]}, \cdots, \bm{u}^{[J]}$ from \eqref{eq:spec4}, the marginal distribution of $\bm{f}$ then reduces to a zero-mean multivariate normal distribution, with covariance matrix:
% \begin{equation}
% &p(\bm{y} |\bm{f}) = \w{N}(\bm{y}| \bm{f}, \sigma^2\bm{I}) \\
% &p(\bm{f}| \bm{u}_1 \dots \bm{u}_J, \bm{X}) = \w{N}(\bm{f}| \bs{\Lambda} \sum_{j=1}^J \bm{A}_j\bs{\Lambda}_j^{-1} \bs{\mu}_j, \bs{\Lambda}) \\
% & p(\bm{u}_j| \bm{Z}_j) = \w{N}(\bm{u}_j| 0, \bm{R}_j)
% \end{equation}
% This model can be interpreted as using $J$ independent GP experts to parametrize the marginal covariance of $\bm{f}$
\begin{equation}
\text{Cov}(\bm{f}) = \bs{\Lambda} + \sum_{j=1}^J \bs{\Lambda} \bm{A}_j \bs{\Lambda}_j^{-1} \bm{K}_{n,m_j}^{[j]} \bm{R}_j^{-1} \bm{K}_{m_j,n}^{[j]}\bs{\Lambda}_j^{-1} \bm{A}_j  \bs{\Lambda}.
\label{eq:cov}
\end{equation}
This thus shows how the employed generative model aggregates the $J$ sparse GP experts for parametrizing the marginal covariance of $\bm{f}$. We show next in Section \ref{sec:var} that, while the covariance matrix in \eqref{eq:cov} and its corresponding precision matrix are dense, we can bypass direct evaluations of its inverse and determinant for model training and prediction, via a carefully-constructed variational inference procedure. 

Here, the choice of the function $\alpha_j(\bm{x})$ plays an important role in how the sparse GP experts are aggregated for global prediction. In our implementation, we used the softmax parametrization:
\begin{equation}
\alpha_j(\bm{x}) = \frac{\exp\left\{-T_j \lambda_j(\bm{x})^c \right\}}{\sum_{k=1}^J\exp\left\{-T_k \lambda_k(\bm{x})^c\right\} }, \quad j = 1, \cdots, J,
\end{equation}
where $\lambda_j(\bm{x})$ is the posterior variance of the $j$-th sparse GP expert at a new point $\bm{x}$, $T_1, \cdots, T_J$ are positive temperature parameters, and $c>0$ is a tuneable hyperparameter. Thus, if the $j$-th sparse GP expert is more certain of its prediction at point $\bm{x}$, i.e., it has lower posterior variance $\lambda_j(\bm{x})$, it will then have greater influence in the aggregate predictor, which is as desired. The parameters $T_1, \cdots, T_J$ and $c$ then control how quickly this expert influence decays with expert uncertainty. In our implementation, these parameters are optimized via variational inference (see Section \ref{sec:var}), to allow for a flexible and efficient calibration of local expert influence from data. The nugget parameters in $\bm{D}_j$ (see \eqref{eq:spec4}) also serve an important purpose: they control the local influence of each sparse GP expert; as $d_{l,j}$ increases, its corresponding pseudo-observation $u^{[j]}_l$ will have less effect on the posterior predictive distribution of $f(\bm{x})$ \citep{stein1999interpolation}. These nugget parameters will again be optimized via variational inference (see Section \ref{sec:var}).

% In our implementation later, the parameters of the soft max function,$\{ \{ T_j\}_{j=1}^J, c\}$, can be efficiently optimized via the variational inference approach presented next, thus enabling a flexible calibration of local expert influence from data.

% Thus, each $d_{k,j}$ can be interpreted as controlling the local influence of the latent expert variable $u_{k, j}$.

An appealing property of the ProSpar-GP model \eqref{eq:spec1}-\eqref{eq:spec4} is that, while each sparse GP expert focuses primarily on modeling \textit{local} activity, the training of its inducing points and pseudo-observations makes use of the \textit{full} dataset $\bm{y}$, instead of just a small local partition. As such, each expert can leverage \textit{global} information over the full prediction space (if needed) to improve local fits. This global property of experts addresses the aforementioned limitation of some existing local GP methods (e.g., the HgPOE \citep{cohen_heal}), which may discard global information in local expert training, and thus may be less effective for prediction given a computational budget. To contrast, the ProSpar-GP can capture both local and global features via the use of sparse GP experts equipped with localized inducing points; we will explore this later in Section \ref{sec:gl}.

% This allows us to avoid learning partitions of the data to model local non-stationarity -- a important advantage over existing local GP expert methods \citep{park2018patchwork,cohen_heal}. Furthermore, each expert may contribute globally to predictions depending on the expert length-scale and inducing point locations.

\mycomment{\cmtS{need here? or later?} The form of the final density on $\bm{f}$ sheds light on why we chose the FITC formulation as opposed to the much more popular SVGP approach. POE weights require inverting the latent function's conditional covariance matrix. In the SVGP stochastic model this covariance matrix is dense, which would require scale $\w{O}(n^3)$. The FITC assumption of conditional independence maintains a diagonal covariance matrix that allows $\w{O}(n)$ inversion.}

\subsection{Variational inference and prediction}
\label{sec:var}
% hich is necessary for huge data sets, prevents use of non-gaussian likelihoods, 
With the hierarchical model \eqref{eq:spec1}-\eqref{eq:spec4} in hand, we now tackle the problem of posterior inference on the pseudo-observations $\bm{u} := \{\bm{u}^{[j]}\}_{j=1}^J$, as well as estimation of its inducing points ${\{\bm{z}_l^{[j]}\}_{l=1}^{m_j}}_{j=1}^J$, associated kernel length-scale parameters $\{\bs{\theta}^{[j]}\}_{j=1}^J$ and other model parameters. One approach might be to directly sample the posterior distribution of pseudo-observations $\bm{u}$ (which can be obtained in closed-form), with model parameters optimized via the maximization of the closed-form marginal likelihood of $\bm{y}$. This ``direct'' approach can, however, be computationally prohibitive with massive datasets, as it requires the inverse computation and matrix multiplication of the dense $(\sum_{j=1}^J m_j) \times (\sum_{j=1}^J m_j)$ covariance matrix \eqref{eq:cov}, which incur $\w{O}\left( (\sum_{j=1}^J m_j)^3\right)$ and $\w{O}\left(n(\sum_{j=1}^J m_j)^2\right)$ complexity for each evaluation of the marginal likelihood, respectively. We thus adopt the following variational inference approach for efficient training of the ProSpar-GP, leveraging the use of mini-batching and GPU acceleration for further scalability.

% and are also difficult to mini-batch for model training.

 % We will instead define a variational family over $\bm{u}$ and find the member that minimizes the kl divergence to the true posterior. For computational efficiency we will assume that the the variational posterior

Our variational inference procedure \citep{blei2017variational} proceeds by first defining an appropriate variational family for approximating the posterior, then optimizing for the distribution in this family that minimizes the Kullback-Liebler (KL) divergence from the desired posterior. We consider the following factorized variational family for approximating the posterior distribution of the pseudo-observations $\mathbf{u}$:
\begin{equation}
    p(\bm{u} | \bm{y}) \approx q(\bm{u}) = \prod_{j=1}^J \phi(\bm{u}^{[j]}; \bm{w}^{[j]}_{\bm{u}} , \bs{\Sigma}^{[j]}_{\bm{u}}),
    \label{eq:vari}
\end{equation}
where $q(\cdot)$ denotes the variational density, and $\{\bm{w}^{[j]}_{\bm{u}}\}_{j=1}^J$ and $\{\bs{\Sigma}^{[j]}_{\bm{u}})\}_{j=1}^J$ are variational parameters. This factorized form yields an easily-computable evidence lower bound (see below) that can be efficiently optimized for the above variational parameters. In later numerical experiments, this factorized form appears to be flexible enough for good predictive performance; one can, however, employ alternate variational forms guided by available prior knowledge, and proceed in a similar fashion below.

% however, one can employ alternate variational forms as suggested by prior knowledge and progress in a similar fashion

With the variational family \eqref{eq:vari}, we can derive the following evidence lower bound (ELBO; \cite{blei2017variational}), which lower bounds the log-marginal likelihood $\log p(\bm{y})$ of the observed data $\mathbf{y}$:
\begin{align}
\begin{split}
\log p(\bm{y}) &= \log\left( \int \frac{p(\bm{y}|\bm{u}) p(\bm{u})}{q(\bm{u})} q(\bm{u}) d\bm{u}\right)\\
& \geq \int q(\bm{u}) \log\left( \frac{p(\bm{y}|\bm{u}) p(\bm{u})}{q(\bm{u})} \right) d\bm{u} \\
 &= \sum_{i=1}^n \left\{ \log \left[\phi \left( y_i; \lambda(\bm{x}_i) \sum_{j=1}^J \frac{\alpha_j(\bm{x}_i)}{\lambda_j(\bm{x}_i) } \bm{k}_{m_j}^{[j]}(\bm{x}_i)^T
    \bm{R}_j^{-1}\bm{w}_{\bm{u}}^{[j]}, \sigma^2 \right) \right] \right. \\
    & \left. \hspace{0.05\textwidth} + \sum_{j=1}^J \frac{ \lambda^2(\bm{x}_i) \left( \frac{\alpha_j(\bm{x}_i)}{\lambda_j(\bm{x}_i) } \right)^2 \bm{k}_{m_j}^{[j]}(\bm{x}_i)^T \bm{R}_j^{-1} \bs{\Sigma}_{\bm{u}}^{[j]} \bm{R}_j^{-1} \bm{k}_{m_j}^{[j]}(\bm{x}_i)}{\lambda^2(\bm{x}_i)  + \sigma^2} \right\}
    - \text{KL}\{q(\bm{u}) || p(\bm{u})\},
\label{eq:elbo}
\end{split}
\end{align}
where $\text{KL}(\cdot||\cdot)$ denotes the KL divergence. The right side of \eqref{eq:elbo} provides the ELBO, which is then maximized with respect to the variational parameters $\{\bm{w}^{[j]}_{\bm{u}}\}_{j=1}^J$ and $\{\bs{\Sigma}^{[j]}_{\bm{u}})\}_{j=1}^J$; this is equivalent to minimizing the KL divergence between the true posterior and its variational approximation (see \citep{blei2017variational}). One can further use this ELBO for fitting the remaining ProSpar-GP model parameters, e.g., its inducing points ${\{\bm{z}_l^{[j]}\}_{l=1}^{m_j}}_{j=1}^J$, kernel length-scale parameters $\{\bs{\theta}^{[j]}\}_{j=1}^J$, softmax weighting parameters $\{ T_j\}_{j=1}^J$ and $c$, and the nugget parameters in $\{\bm{D}_j\}_{j=1}^J$. Specifically, these parameters are set to maximize the ELBO; this can be viewed as an empirical Bayes estimation of such parameters, as the ELBO lower bounds the log-marginal likelihood \cite{carlin1997bayes}. Parameter estimation via the ELBO (or marginal likelihood) also helps regularize against over-fitting \citep{lotfi2022bayesian, rasmussen_occam_2000}, which is important for estimating expert-specific length-scale parameters in our model.
% , such an procedure can be interpreted as an empirical Bayes estimation and is widely implemented as Type-II maximum likelihood optimization in machine learning literature 
 % The ELBO can then be maximized with respect to the variational parameters $\{\bm{w}^{[j]}_{\bm{u}}\}_{j=1}^J$ and $\{\bs{\Sigma}^{[j]}_{\bm{u}})\}_{j=1}^J$, to find the closest variational approximation to the desired posterior $[\bm{u}|\bm{y}]$ in KL-divergence. 

There are two key advantages in using this variational approach for parameter estimation over the aforementioned direct maximization of the marginal likelihood. First, note that the ELBO \eqref{eq:elbo} sums over the $n \gg 1$ observations in the big data. This allows for unbiased gradient estimates of the ELBO via Monte Carlo approximation using random subsamples (or \textit{mini-batching}) of the large dataset; such estimates can then be directly integrated within mini-batch stochastic gradient descent methods \citep{li2014efficient} for scalable parameter optimization. Similar stochastic variational inference approaches have shown successful results for scaling up standard GPs (see, e.g., \cite{ppgr, hensman2013gaussian}), and appears to work well for our model later. Second, each evaluation of the ELBO \eqref{eq:elbo} requires only operations on the $J$ smaller $m_j \times m_j$ matrices, rather than the operations on the large $\sum_{j=1} m_j \times \sum_{j=1} m_j$ covariance matrix required for the direct marginal likelihood. This greatly reduces computation for parameter estimation, leading to significant speed-ups for model training with massive data. A full analysis of computational complexity is provided later in Section \ref{sec:comp}. The above variational approach for parameter estimation (in particular, of local inducing points and length-scales of each expert) thus provides a \textit{scalable} and \textit{data-adaptive} way of identifying local regions of non-stationarity from massive datasets.

% Here, $\lambda^{i} = , \lambda^{i}, A^{i}_j$ are the $i$-th diagonal element of the matrices $\bs{\Lambda}, \bs{\Lambda}_j, \bm{A}_j$ respectively. 

% $\Lambda^i$, $\Lambda_j^i$ and $A_j^i$ refer to the ith diagonal element of the matrices $\bs{\Lambda}, \bs{\Lambda}_j, \bm{A}_j$ respectively. This bound is very similar to the variational FITC bound given by \cite{ppgr} in their formulation of parametric gaussian Process regressors. The bound allows us to accelerate model training by estimating the elbo from sampled mini-batches of observations. Detailed derivation of this bound can be found in the supplementary material.

Finally, with optimized variational parameters $(\{\bm{w}^{[j]}_{\bm{u}}\}_{j=1}^J,\{\bs{\Sigma}^{[j]}_{\bm{u}})\}_{j=1}^J)$ and estimated model parameters, we can then approximate the desired posterior predictive distribution $[f(\bm{x}_{\rm new})|\bm{y}]$ at a new point $\bm{x}_{\rm new}$, by marginalizing against the variational form \eqref{eq:vari}:
\begin{align}
\begin{split}
 [f(\bm{x}_{\rm new})|\bm{y}] = \int [f(\bm{x}_{\rm new})|\bm{u}] [\bm{u}|\bm{y}] \; d\bm{u} &\approx \int [f(\bm{x}_{\rm new})|\bm{u}] \prod_{j=1}^J \phi(\bm{u}_j; \bm{w}^{[j]}_{\bm{u}} , \bs{\Sigma}^{[j]}_{\bm{u}}) \; d\bm{u}\\
 &= \phi(f(\bm{x}_{\rm new}); \mu_{\rm new}, \sigma^2_{\rm new}).
 \label{eq:marg}
\end{split}
\end{align}
Here, $\mu_{\rm new}$ and $\sigma^2_{\rm new}$ have the closed-form expressions:
\begin{align}
\begin{split}
\mu_{\rm new} &= \lambda(\bm{x}_{\rm new}) \sum_{j=1}^J \frac{\alpha_j(\bm{x}_{\rm new})}{\lambda_j(\bm{x}_{\rm new})} 
\bm{k}_{m_j}^{[j]}(\bm{x}_{\rm new})^T \bm{R}_j^{-1} \bm{w}^{[j]}_\bm{u},\\
\sigma^2_{\rm new} &= \lambda(\bm{x}_{\rm new}) +  \sum_{j=1}^J \lambda_j^2(\bm{x}_{\rm new}) \bm{k}_{m_j}^{[j]}(\bm{x}_{\rm new})^T \bm{R}_j^{-1} \bs{\Sigma}_{\bm{u}}^{[j]}  \bm{R}_j^{-1}  \bm{k}_{m_j}^{[j]}(\bm{x}_{\rm new}).
\label{eq:predclosed}
\end{split}
\end{align}
Thus, with optimized variational and model parameters, one can then use such closed-form expressions to efficiently sample from the (approximated) posterior predictive distribution of $f(\bm{x}_{\rm new})$.

% Together with the earlier variational inference procedure, this provides a scalable end-to-end framework for GP modeling with massive non-stationary data.

% \subsection{Mini-batching and GPU acceleration}

\section{Properties of the ProSpar-GP}
\label{sec:prop}

We now explore key appealing properties of the ProSpar-GP and how they compare to existing methods. This includes an investigation of runtime and memory complexity, Kolmogorov consistency, and global-local modeling properties.

% \subsection{Advantages}

\subsection{Computational and memory complexity}
\label{sec:comp}

Recall that the key computational bottleneck for standard GP modeling is the required $\mathcal{O}(n^3)$ work and $\mathcal{O}(n^2)$ memory, where $n$ is the sample size of the large training dataset. In the following, we investigate the runtime and memory complexities of the ProSpar-GP, and compare it with the existing state-of-the-art in terms of addressing this bottleneck.

 For \textit{model training} of the ProSpar-GP, its complexity is dominated by the computational cost in evaluating the mini-batched ELBO objective, i.e., the unbiased Monte Carlo approximation of \eqref{eq:elbo} using random subsamples of size $B \ll n$ from the large training dataset. Each evaluation of this mini-batch ELBO requires Cholesky decompositions of $J$ matrices, each of size $m_j \times m_j$, where $m_j$ is the number of inducing points for the $j$-th expert. With this, one can show that each mini-batch ELBO evaluation requires a runtime complexity of $\w{O}\left( B \sum_{j=1}^J m_j^2 + \sum_{j=1}^J m_j^3 \right)$ and a memory complexity of $\w{O}\left( \sum_{j=1}^J m_j^2\right)$. Using the trained model, \textit{predictions} from the ProSpar-GP via the posterior predictive distribution \eqref{eq:marg} requires a runtime of $\w{O}\left( \sum_{j=1}^J m_j^2 \right)$ and memory complexity $\w{O}\left( \sum_{j=1}^J m_j^2\right)$, assuming the $J$ Cholesky decompositions from model training are re-used for prediction. To contrast, standard stochastic variational inducing point approaches \citep{tt_inducing} require a runtime and memory cost of $\w{O}\left( B m^2 + m^3 \right)$ and $\w{O}\left( m^2\right)$ respectively, where $m$ is the number of total inducing points used for approximation. Viewed this way, the ProSpar-GP allows for quicker computation over existing inducing point approaches given the \textit{same} number of inducing points $m$, particularly when $m_j$ (the number of inducing points for an expert $j$ in the ProSpar-GP) is much less than $m = \sum_{j=1}^J m_j$ (the total number of inducing points). Moreover, in the presence of non-stationarity, further computational gains can be realized as standard inducing point methods require considerably more inducing points over the ProSpar-GP for accurate approximation (see Section \ref{sec:sgp}).

  % to use the same number of total inducing points with far fewer computational resources

% Define mini-batch size of $B$ using $J$ experts each with $m_j$ inducing points. Under these settings, each iteration of the ProSpar-GP objective has $\w{O}( \sum_{j=1}^J B m_j^2 + \sum_{j=1}^J m_j^3)$ runtime complexity and $\w{O}( \sum_{j=1}^j m_j^2)$ space complexity.

The computational advantage of the ProSpar-GP extends beyond the above runtime analysis. A quick inspection of \eqref{eq:elbo} shows that a significant portion of the ELBO evaluations (and subsequent operations for computing the predictive distribution \eqref{eq:marg}) involves matrix multiplications, which  can be greatly sped up via standard off-the-shelf GPU software, e.g., PyTorch \citep{paszke2019pytorch}, JAX \citep{black_box, jax2018github} or TensorFlow \citep{tensorflow2015-whitepaper}. This straight-forward integration of GPU acceleration within the ProSpar-GP is a potential advantage over Vecchia-approximation methods, which require either populating individual elements of a sparse Cholesky matrix, or performing $n$ Cholesky decompositions (each with complexity $\w{O}(q^3)$, where $q$ is the number of nearest neighbors used). Neither of these operations easily exploit GPU acceleration \cite{black_box}, and may require an expert GPU programmer to realize tangible computational gains. Our approach does not suffer as much from this problem, as it only requires $J \ll n$ Cholesky decompositions. Further, with multiple GPUs, much of the required computation for ProSpar-GP model training and prediction can easily parallelized by dividing the experts among GPU units. Our later numerical experiments make use of a single GPU for computation.

\subsection{Kolmogorov consistency}

An appealing property of the ProSpar-GP is that its Kolmogorov consistency \citep{tao2011introduction}, in that the generative distribution in  \eqref{eq:spec1}-\eqref{eq:spec4} defines a valid stochastic process over $\mathbb{R}^d$; such consistency is critical for stable model training using variational inference. This is formally stated below:
\begin{prop}
Consider the marginal density specification \eqref{eq:spec1}-\eqref{eq:spec4} from the ProSpar-GP on the latent function values $\bm{f}$ given a finite set of input points $\bm{x}_1, \cdots, \bm{x}_n \in \mathcal{X}$. It follows that:
\begin{enumerate}
    \item This marginal density (denoted $p(\bm{f})$) is permutation-invariant, i.e., for any choice of $\bm{x}_1, \dots, \bm{x}_n$ and any permutation $\pi$ of $\{1, \cdots, n\}$, we have:
    \begin{equation}
        p(f(\bm{x}_1), \cdots, f(\bm{x}_{n})) = p(f(\bm{x}_{\pi(1)}), \cdots, f(\bm{x}_{\pi(n)})).
        \label{eq:kol1}
    \end{equation}
    \item For any two sets of latent function realizations $\bm{f}_1$ and $\bm{f}_2$, we have:
    \begin{equation}
    p(\bm{f}_1) = \int p(\bm{f}_1, \bm{f}_2) d\bm{f}_2.
    \label{eq:kol2}
    \end{equation}
\end{enumerate}
Thus, by the Kolmogorov extension theorem (Theorem 2.4.3 of \citep{tao2011introduction}), there exists a valid stochastic process on $\mathbb{R}^d$ satisfying the ProSpar-GP marginal distribution $p(\bm{f})$.
\label{prop:cons}
\end{prop}
\noindent The proof of this proposition is provided in Supplementary Materials. The key idea is to leverage the so-called Kolmogorov extension theorem (Theorem 2.4.3 of \citep{tao2011introduction}) to guarantee the existence of a (infinite-dimensional) stochastic process that satisfies a prescribed specification of its finite-dimensional marginal distribution. This consistency argument is crucial for ensuring that inference is performed under a valid probabilistic model \citep{samo2016string}.

We note that many existing product-of-expert approaches for GPs (e.g., \citep{tresp, distributed, cohen_heal, cpoe, bcm, rbcm}) may not satisfy Condition 2 in the above proposition (see \citep{samo2016string}).  This inconsistency can greatly limit the model flexibility and stability of such approaches, and may explain why these methods tend to perform better with identical expert length-scale parameters as a safeguard against over-fitting \cite{cohen_heal, distributed}. To contrast, the ProSpar-GP relies on a valid generative stochastic process (Proposition \ref{prop:cons}) when making inference on model hyperparameters; this then improves the stability of parameter inference without need for model simplifications, which can be detrimental for non-stationary modeling. Such consistency is also critical for ensuring stability in mini-batch optimization; if violated, the marginal likelihood to optimize may change between mini-batches, which is highly undesirable.

% In addition, the lack of an underlying stochastic process forced severely limited the flexibility of POE methods.

% for downstream analysis tasks such as computer model calibration \citep{hagan_calibration}

\subsection{Global-local modeling}
\label{sec:gl}
Finally, another appealing property of the ProSpar-GP is that, by allowing local sparse GP experts access to the full (global) dataset $\bm{y}$, it inherits both the desired computational efficiency of local-expert models and the predictive power of global models. This is in contrast with many existing local GP expert methods, which may sacrifice the latter for scalability with big data. This ``global-local'' property can facilitate scalable prediction given a computational budget, particularly for non-stationary surfaces with global trends but local regions of variability. To explore this further, we return to our earlier 1-d example from Section \ref{sec:sgp}, which featured a non-stationary function with distinct local and global features. Such non-stationarity posed a challenge for standard sparse GP methods, as seen earlier in Figure \ref{fig:illustrate}. We now apply the proposed ProSpar-GP with $J=9$ experts, each equipped with $m_j = 5$ inducing points, where inducing point locations and GP length-scales (along with variational parameters) are optimized via the ELBO \eqref{eq:elbo}. 

\begin{figure}[!t]
\centering
\subfloat[]{%
  \includegraphics[width=7cm]{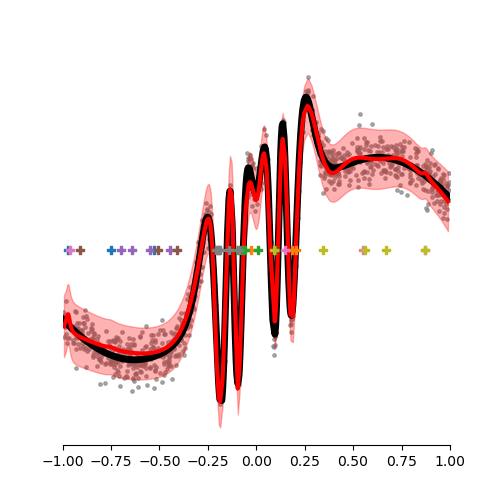}%
  \label{fig:illustrate:mine}%
}
\subfloat[]{%
  \includegraphics[width=7cm]{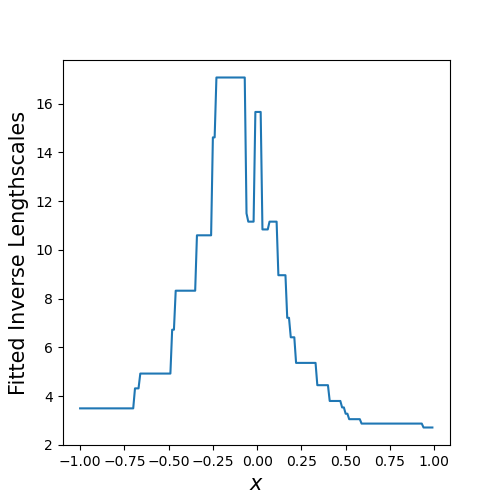}%
  \label{fig:illustrate:lengthscales}%
} 
\caption{(a, left) Visualizing the fitted ProSpar-GP with $J=9$ experts and $m_j = 5$ inducing points on the earlier 1-d example. The colored crosses mark inducing point locations, with each color representing a different sparse GP expert. (b, right) Visualizing the fitted inverse length-scale parameters for the ProSpar-GP experts via a 10-nearest-neighbor smoother.}
\label{fig:illustrate}
\end{figure}

Figure \ref{fig:illustrate:mine} shows the corresponding ProSpar-GP fit, along with its optimized inducing points (colored by expert). We highlight two interesting observations. First, note that the optimized inducing points are largely situated within the desired middle interval $[-0.5,0.5]$, where there is high fluctuation in $f$. This addresses the earlier limitation of standard sparse GPs from Section \ref{sec:sgp} (see Figure \ref{fig:illustrate:optimize}), where in neglecting non-stationarity, the optimized inducing points can be far from regions of high activity. By optimizing such points within a product-of-experts framework, the ProSpar-GP (Figure \ref{fig:illustrate:mine}) appears to do well at allowing experts to carefully adapt to and model for \textit{local} non-stationary features. Second, we see that the fitted ProSpar-GP experts can indeed learn and model for \textit{global} behavior in $f$. Recall from Figure \ref{fig:illustrate:ls} that, even when inducing points are manually placed within regions of high activity, the resulting sparse GP fit does not capture well the underlying global trend of the function, particularly near the ends of the prediction space. The ProSpar-GP addresses this (Figure \ref{fig:illustrate:mine}) by carefully placing some experts for modeling the underlying global structure, and other experts to model local non-stationary fluctuations. With this global-local modeling property, the fitted ProSpar-GP achieves noticeably improved prediction with better calibrated uncertainties for this non-stationary example.

Figure \ref{fig:illustrate:lengthscales} further investigates this global-local property. The plotted curve visualizes how the fitted inverse length-scale parameters for the $J=9$ ProSpar-GP experts change over the domain, using a 10-nearest-neighbor smoother on the optimized inducing points with fitted inverse length-scales as labels. We see a clear peak in these inverse length-scales within the region of high volatility (near the center), and low inverse length-scales within regions of low volatility (at the end-points). Thus, for this test function with global structure and local fluctuations, the ProSpar-GP captures both properties well via a careful placement of local experts (with small length-scales) and global experts (with large length-scales) over appropriate regions of the domain.

\begin{figure}[!t]
\centering
\subfloat[ProSpar-GP]{%
  \includegraphics[width=0.34\textwidth]{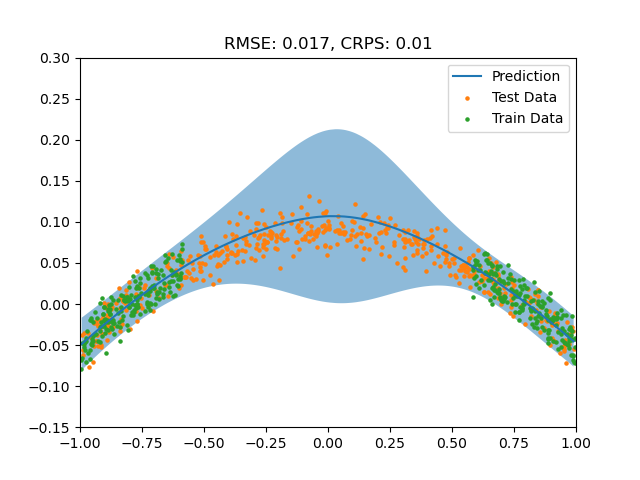}%
  \label{fig:interp:min}%
}
\subfloat[HgPOE]{%
  \includegraphics[width=0.34\textwidth]{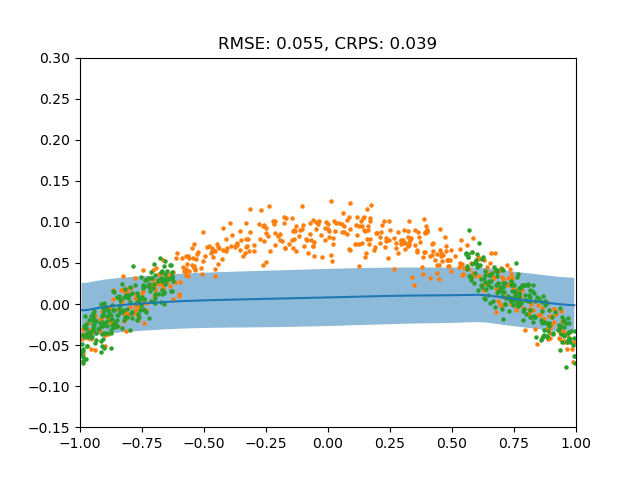}%
  \label{fig:interp:gpoe}%
} 
\subfloat[HLaGP]{%
  \includegraphics[width=0.34\textwidth]{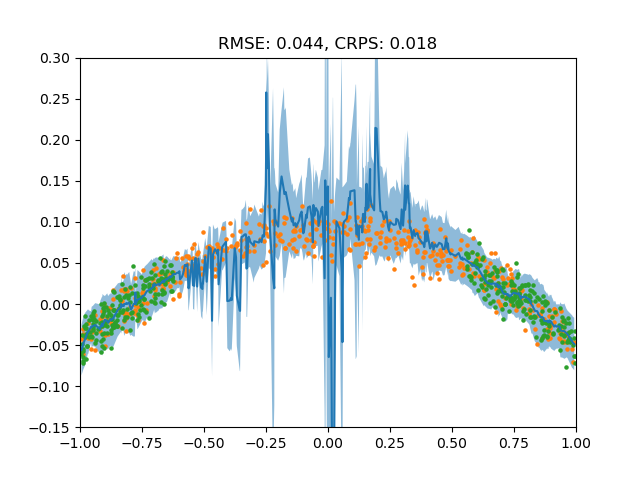}%
  \label{fig:interp:hlagp}%
}

\caption{Visualizing the predictive performance of the ProSpar-GP and two local GP expert methods (HgPOE and HLaGP), for an experiment with two disjoint training datasets (green dots). The posterior mean predictor is shown as the solid blue curve, with its 95\% predictive intervals shaded in blue.}
\label{fig:interp}
\end{figure}

Finally, we consider a more challenging experiment to compare this global modeling property with existing local GP methods. Figure \ref{fig:interp} visualizes this set-up. Using a function with a clear global trend, we generate from it two large disjoint training datasets, with no data in the middle of the input space. We then compare the performance of our method (with $J=20$ experts and $m_j = 20$ inducing points) with two state-of-the-art \textit{local} GP experts methods: the HgPOE \citep{cohen_heal} (with 20 experts) and the HLaGP \citep{sun} (with local designs of size 20). Methods that can best exploit the underlying global structure should thus provide good predictions within the data-sparse middle region. Figures \ref{fig:interp:min}-\ref{fig:interp:hlagp} show the corresponding predictions (with uncertainty) for the three methods. We see that the HgPOE fails to capture this global trend, which is unsurprising since it aggregates GP models on \textit{disjoint} partitions. The HLaGP provides improved global modeling via a careful construction of conditioning sets, but still yields erratic predictions and uncertainties within the data-sparse region. The ProSpar-GP provides the best performance of the compared methods, both visually and in terms of RMSE and CRPS. This thus suggests that the sparse GP experts in our model, each having access to the full (global) dataset, can indeed learn and exploit global trends for better predictive performance. This improved prediction within data-scarce regions becomes increasingly important in \textit{higher} dimensions, when there are larger gaps between training data points; we will see this next in numerical experiments.

% We set ProSpar with 20 experts with 20 inducing points each. GPOE is set with 20 experts allocated with 20 observations each.  HLaGP is given with local design sizes of 20. 

% ProSpar successfully interpolates the structure of the function and expresses reasonable uncertainty in the data-sparse region. The GPOE model completely fails to capture the global dynamics. HLaGP careful construction of conditioning sets allows it to performs better, but it still provides erratic predictions. In particular, We have seen how the distinct sparse experts with localized inducing points allow ProSpar to adapt to local features and non-stationarity. However, each sparse expert in ProSpar is also a global model in that the parameters are optimized with respect to the entire training set and each expert contributes globally to predictions made across the input space. Compared to other local expert models, this characteristic allows the ProSpar to capture global trends. \textcolor{red}{In addition it avoids having to rely of prefixed partitions of data or learning such partitions}

% \subsection{Back to illustrating example}
% \label{sec:localization}
\mycomment{
ProSpar is not inherently a local experts model as each experts affects the entire input space. To localize ProSpar, we can allocate each expert a small number of inducing points initialized in local regions of the input space using an algorithm such as kmeans. This initialization encourages each expert to govern local regions of the input space. Optimization of the inducing point locations via  the ELBO then allows experts to adapt to local features of the data. To illustrate the benefits of this approach we return to the example in \ref{fig:illustrate}. We now apply ProSpar with $J =9$ experts each with  $m = 5$ inducing points. Figure \ref{fig:illustrate:mine} shows the ProSpar fit and inducing point locations represented by the colored crosses. Each color represents an distinct expert with unique kernel hyper-parameters. We see that unlike the standard inducing point fits in \ref{fig:illustrate}, ProSpar is able to adequately place a high density of points in the area of fast variation.

The ProSpar formulation of non-stationarity also allows us to easily interpret the model to identify areas of fast variation. We associate each inducing point with the length-scale of its expert. On a grid defined across the input space, we can then perform to a K-Nearest Neighbor regression on the inducing point inverse lengthscales to obtain a measure of the variation in the region. Figure \ref{fig:illustrate:lengthscales} demonstrates the results of this procedure applied to the ProSpar fit using 10-Nearest Neighbors. We see that there is a clear peak in the averaged inverse length-scale of inducing points in the central area of the fast variation indicating that ProSpar adequately learned the local features.
}

\section{Numerical Experiments}
\label{sec:exp}

We now explore the ProSpar-GP in a suite of numerical experiments that feature large training datasets on non-stationary response surfaces, and compare its performance with existing state-of-the-art methods. We first outline the compared methods and their implementation, then investigate its performance in terms of point and probabilistic predictions.

\subsection{Set-up and benchmark methods}
\label{sec:comp_imp}

Following Section \ref{sec:recent}, we choose several benchmarks from state-of-the art methods using inducing points, Vecchia approximations, deep GPs and local GP expert methods. Hyperparameter settings for each approach (detailed below) are either fixed at recommended settings from the provided paper and/or package, or are chosen to provide comparable computing power between all methods. The compared methods are listed below:

% Implementing each competing method requires choosing hyper-parameter values that control the quality of the approximation at the expense of computation. For each competitor, we choose hyper-parameter values to match or exceed the approximation quality compared to the implementations in their original papers.

% This section compares the performance of ProSpar to current state of the art gaussian process emulation and approximation methods.

\begin{itemize}[leftmargin=*]
    \item \textbf{Stochastic Variational Gaussian Process} (SVGP; \cite{tt_inducing}): The SVGP is a widely-adopted inducing point approach for scalable GP regression \citep{hensman2013gaussian, bauer}. We used the implementation in the Python package \texttt{GpyTorch} \citep{black_box}, with the Mat\'ern-5/2 kernel, $m = 1024$ inducing points, and a mini-batch size of $B=1024$. Such settings are widely used in the literature.

    \item \textbf{Vecchia-approximated Deep Gaussian Process} (VDGP; \cite{sauer2022vecchiaapproximated}): The VDGP is a recent development on Vecchia-approximated GPs. We used the implementation in the \textsc{R} package \texttt{deepgp} \citep{sauer2022vecchiaapproximated} with the recommended two layers. Each layer makes use of a GP with the Mat\'ern-5/2 kernel \citep{stein1999interpolation}. Here, we followed the recommended setting of $q = 25$ nearest neighbors and 1000 MCMC iterations; while performance may improve with additional MCMC iterations, even 1000 iterations can take more than a day to run for some experiments.
    
    % Input data is pre-scaled using length-scales obtained from a vecchia approximated GP with $m = 25$. 

    \item \textbf{Scaled Vecchia-approximated Gaussian process} (SVecGP; \cite{scaledvecchia}): The SVecGP is a recent Vecchia GP method that constructs conditioning sets via a scaled Euclidean distance \cite{Katzfuss_2021}. Model hyperparameters are estimated via a randomly subsampled dataset of $10,000$ points. Here, we used the provided code from the paper, with the Mat\'ern-5/2 kernel and $q = 25$ and $q=150$ nearest neighbors; these will be denoted as SVecGP(25) and SVecGP(150), respectively.
    
    \item \textbf{Hybrid Local Approximate Gaussian Process} (HLaGP; \citep{sun}): The HLaGP is a recent local GP method that hybridizes global and local information to make predictions for individual test points. We used the implementation from the \textsc{R} package \texttt{lagp}  \cite{lagp}, with local designs of size $50$ constructed via the Active Learning Cohn criterion. Here, local GP experts are equipped with the squared-exponential kernel with length-scale parameters estimated from data.
    % \footnote{\texttt{lagp} only supports squared-exponential kernel} 
    % \textcolor{red}{The approach allows non-stationarity in the predictions}.

    \item \textbf{Healed generalized Product of Experts} (HgPOE; \cite{cohen_heal}): The HgPOE (as described in Section \ref{sec:recent}) is implemented from the provided code in the paper, using softmax weights with the recommended temperature parameter of $T = 100$. Here, each expert employs a separate squared-exponential kernel, and is assigned approximately $500$ data points, thus yielding $J \approx n/500$ experts. These data points were assigned to each expert via k-means clustering of the input points, as recommended in \cite{cohen_heal}. We have tried reducing the number of experts in implementation, but this resulted in numerical instabilities and much higher computational costs.
    % Attempts to assign each expert more data points resulted in much higher computational cost and numerical instabilities. 
    \item \textbf{ProSpar-GP}: Our method is implemented using separate squared-exponential kernels (with distinct length-scale parameters) for each sparse GP expert. Here, $J = 200$ experts are used, each with $m_j = 25$ inducing points. The locations of these inducing points, as well as model hyperparameters, are optimized via variational inference (see Section \ref{sec:var}). We made use of the Adam algorithm \citep{kingma2017adam} for optimizing the ELBO \eqref{eq:elbo}, along with a mini-batch size of $B=1024$.
\end{itemize}
\noindent For methods involving local modeling (HLaGP, HgPOE and ProSpar-GP), we found that the squared-exponential kernel works slightly better than the Mat\'ern kernel. This may be because the dangers of global over-smoothing is not as pronounced for local models, which may further benefit from the regularization of a smoother kernel. This observation is consistent with the implementation of local expert models in the literature \citep{tresp, cohen_heal, lagp, samo2016string}, All methods are then compared on two metrics: the RMSE, which measures quality of point predictions, and the continuous ranked probability score (CRPS; \cite{gneiting2007strictly}), which measures quality of probabilistic predictions.

\subsection{Results}

\begin{figure}[!t]
    \centering
    \includegraphics[width = \textwidth]{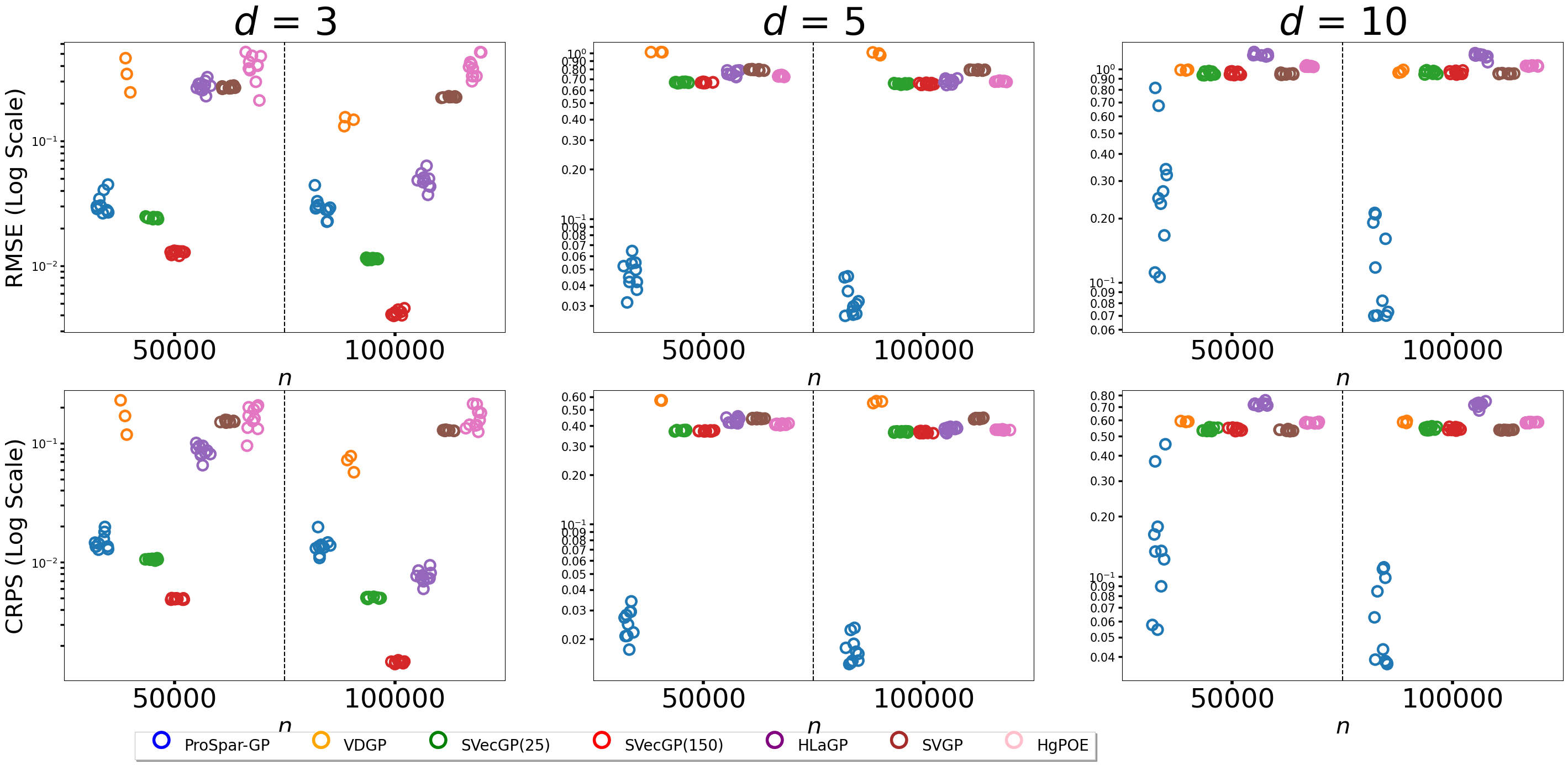}
    \caption{Predictive metrics (top: RMSE in log-scale; bottom: CRPS in log-scale) of the compared methods in the Michalewicz experiment, for different dimensions $d$ and sample sizes $n$.}
    \label{fig:michal_results}
\end{figure}

\begin{table}[!t]
    \centering
\begin{tabular}{cccccccc}
\toprule
\multicolumn{8}{c}{\textbf{Michalewicz} ($n=50,000$ samples)} \vspace{0.1cm}\\
$d$ & ProSpar-GP & VDGP & SVecGP($25$) & SVecGP($150$) & HLaGP & SVGP & HgPOE \\
\midrule
3 & 6.65 & 544.09 & 0.44 & 22.95 & 64.16 & 15.02 & 3.01 \\
5 & 7.43 & 981.60 & 0.55 & 22.54 & 57.83 & 14.73 & 3.19 \\
10 & 10.43 & 1658.27 & 1.04 & 44.50 & 46.10 & 14.69 & 3.30 \\
\toprule
\multicolumn{8}{c}{\textbf{Michalewicz} ($n=100,000$ samples)}\vspace{0.1cm}\\
$d$ & ProSpar-GP & VDGP & SVecGP($25$) & SVecGP($150$) & HLaGP & SVGP & HgPOE \\
\midrule
3 & 13.29 & 1028.09 & 0.45 & 23.01 & 91.96 & 29.33 & 5.89 \\
5 & 23.98 & 2012.85 & 0.54 & 22.71 & 43.61 & 29.10 & 7.82 \\
10 & 30.18 & 2919.13 & 1.09 & 46.38 & 53.30 & 29.03 & 6.07 \\
\bottomrule
\end{tabular}
    \caption{Total CPU hours (in minutes) required for model training and prediction of the compared methods in the Michalewicz experiment, for different dimensions $d$ and sample sizes $n$.}
    \label{tab:michal}
\end{table}

We first explore the performance of these methods for the aforementioned Michalewicz function \citep{MARREL2009742, sun, sauer2022vecchiaapproximated}, a common test function for surrogate modeling and uncertainty quantification \citep{simulationlib}. As is clear from Figure \ref{fig:michal:michal}, this function is highly non-stationary, with steep valleys interspersed with flat areas of low variation. We investigate this function in $d = 3$, $5$ and $10$ dimensions, using $n=50,000$ and $100,000$ training samples drawn from a Latin hypercube design \citep{lhc} with noise variance $\gamma^2 = 10^{-8}$.  We evaluate prediction accuracy on $25,000$ test points obtained in a similar fashion. This simulation is replicated 10 times for each method, except for the VDGP, which is replicated 3 times due to its high computation cost (see Section \ref{sec:comp_imp}).

Figure \ref{fig:michal_results} shows the RMSE and CRPS in the Michalewicz experiments, for different dimensions $d$ and sample sizes $n$. In low dimensions ($d=3$), we see that the local-neighbor methods (SVecGP and HLaGP) yield the best performance for both point and probabilistic predictions, the proposed ProSpar-GP providing comparable (but slightly worse) performance, and the HgPOE and SVGP yielding mediocre performance. The excellent performance of existing local-neighbor methods is not surprising: with closer vicinity of points in \textit{low} dimensions, the choice of conditioning sets is less important. In higher dimensions ($d=5$ and $d=10$), the same local-neighbor methods (SVecGP and HLaGP) quickly deteriorate in performance for both RMSE and CRPS (as was noted from Section \ref{sec:recent}), with similarly mediocre performance for the HgPOE\footnote{Here, we tried to improve predictions for the HgPOE by varying the number of experts (and thereby the number of points per expert), but this did not appear to improve performance.} and SVGP. The ProSpar-GP provides considerably improved predictions over existing methods in this challenging higher-dimensional setting, yielding much lower RMSE and CRPS. The improvement of the ProSpar-GP over the inducing points SVGP approach is also worth noting, where the latter (see Section \ref{sec:recent}) can yield poor approximations in non-stationary settings. Here, the proposed approach appears to identify important non-stationary structure of the Michalewicz function within the product-of-experts framework, then leverage this learned local structure for effective prediction, particularly in higher dimensions. 

As different methods are optimized for different computing systems, an ``apples-to-apples'' comparison of computation time may be difficult here. For example, the proposed ProSpar-GP and SVGP leverage speed-ups from GPU acceleration, the VDGP and HLaGP are optimized for parallel computation, while SVecGP and HgPOE are not as easily parallelized. In our implementation, the ProSpar-GP, HgPOE and SVGP are run on a single Nvidia RTX 2080 Ti GPU, the VDGP and HLaGP are run on eight Intel Xeon Gold 6252 CPU cores, and the SVecGP is run on a single such core. With this in mind, however, Table \ref{tab:michal} provides a rough snapshot of computing costs, by reporting the total processing time (in minutes) required for model training and prediction, over different dimensions $d$ and sample sizes $n$. We see that the ProSpar-GP enjoys comparable computing times with the quickest methods, with only the SVecGP(25) and HgPOE being faster. The latter two methods, despite being highly efficient, can yield poor predictions, particularly in higher dimensions. It is worth noting that the ProSpar-GP runs around twice as fast as the inducing points SVGP approach, despite the former using almost five times the number of inducing points. This speed-up highlights the benefit of dividing inducing points over different local experts, which can jointly reduce the computational cost of Cholesky decompositions (see Section \ref{sec:comp}) as well as allow for localized modeling of non-stationary features.

\begin{figure}[!t]
    \centering
    \includegraphics[width = \textwidth]{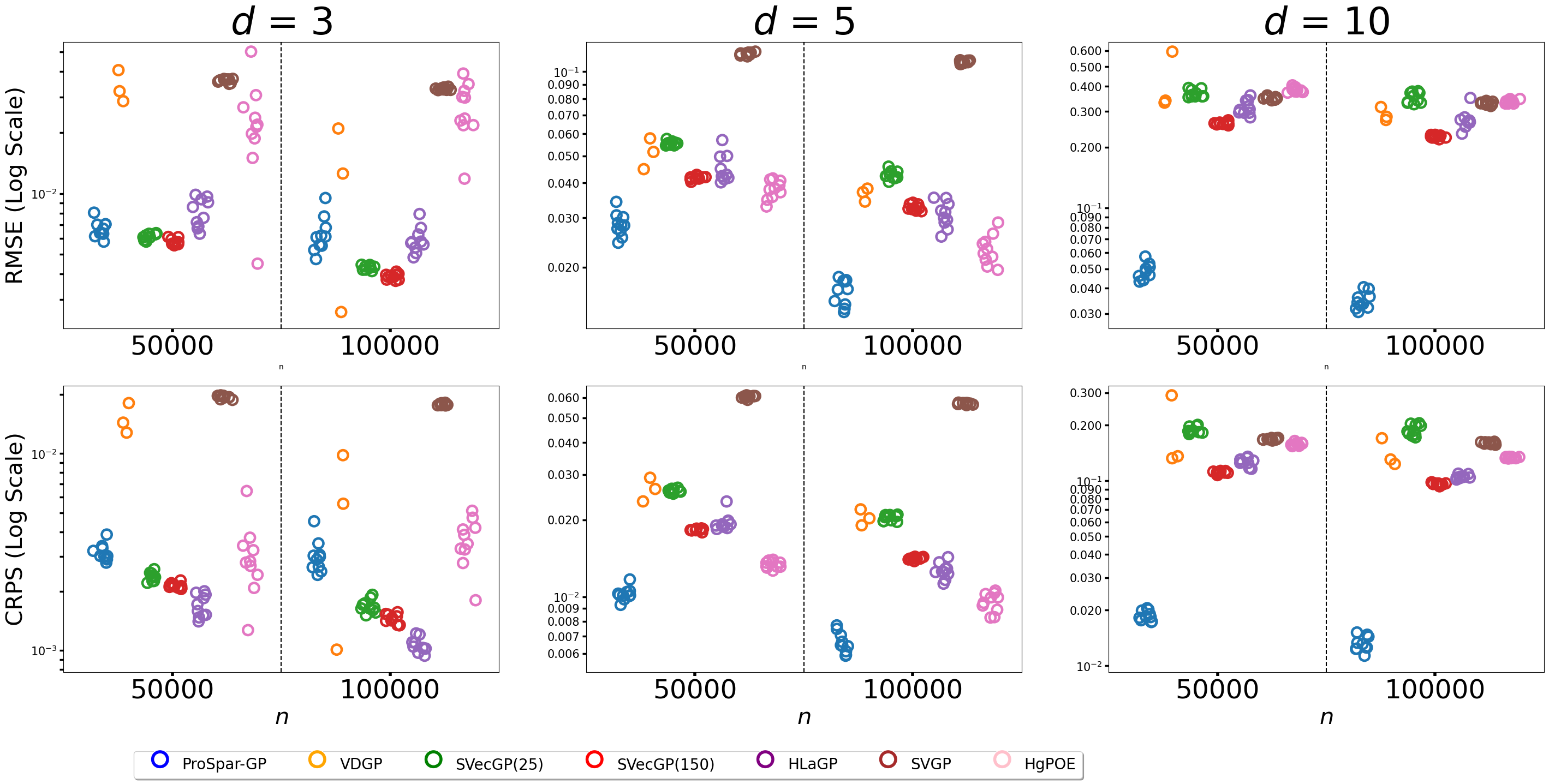}
    \caption{Predictive metrics (top: RMSE in log-scale; bottom: CRPS in log-scale) of the compared methods in the G-function experiment, for different dimensions $d$ and sample sizes $n$.}
    \label{fig:gfunc_results}
\end{figure}

\begin{table}[!t]
    \centering
\begin{tabular}{cccccccc}
\toprule
\multicolumn{8}{c}{\textbf{G-function} ($n=50,000$ samples)} \vspace{0.1cm}\\
$d$ & ProSpar-GP & VDGP & SVecGP($25$) & SVecGP($150$) & HLaGP & SVGP & HgPOE \\
\midrule
3 & 6.65 & 527.96 & 0.29 & 18.11 & 55.53 & 14.83 & 3.34 \\
5 & 7.34 & 1069.63 & 0.36 & 31.02 & 64.10 & 14.84 & 3.08 \\
10 & 9.34 & 2089.28 & 1.05 & 33.45 & 46.10 & 14.71 & 2.78 \\
    \toprule
\multicolumn{8}{c}{\textbf{G-function} ($n=100,000$ samples)} \vspace{0.1cm}\\
$d$ & ProSpar-GP & VDGP & SVecGP($25$) & SVecGP($150$) & HLaGP & SVGP & HgPOE \\
\midrule
3 & 13.34 & 1100.96 & 0.30 & 17.94 & 72.79 & 29.75 & 6.65 \\
5 & 14.56 & 2107.15 & 0.36 & 30.75 & 48.31 & 29.41 & 8.63 \\
10 & 18.61 & 2757.93 & 1.06 & 29.14 & 54.17 & 29.31 & 7.48 \\
\bottomrule
\end{tabular}
    \caption{Total processing time (in minutes) required for model training and prediction of the compared methods in the G-function experiment, for different dimensions $d$ and sample sizes $n$.}
\label{tab:gfunc_runtimes}
\end{table}

Next, we investigate these methods for another common test function in uncertainty quantification: the G-function \citep{simulationlib}. This function again exhibits non-stationary behavior: it has fairly gradual variations near the origin, but abrupt variations towards the edges of the domain. As before, we perform this experiment in $d=3$, $5$ and $10$ dimensions, using $n= 50,000$ and $100,000$ samples drawn from a Latin hypercube design \citep{lhc} with noise variance $\gamma^2 = 10^{-8}$. These simulations are replicated in the same fashion as the earlier experiment.

Figure \ref{fig:gfunc_results} shows the RMSE and CRPS in the G-function experiments for different dimensions $d$ and sample sizes $n$. We observe similar results as before. In low dimensions ($d = 3$), local-neighbor methods (SVecGP and HLaGP) perform the best, with comparable (but slightly worse) predictions from the ProSpar-GP, and mediocre performance for the HgPOE and SVGP. This is again unsurprising, since for local-neighbor methods in \textit{low} dimensions, the precise selection of conditioning sets is less important as observations are closer together. When dimension $d$ increases ($d=5$ and $d=10$), the performance of existing methods again greatly deteriorates, and in this more challenging setting, the ProSpar-GP considerably outperforms its competitors in terms of predictive performance. Table \ref{tab:gfunc_runtimes} summarizes the computing times for model training and prediction, where we again see that the ProSpar-GP enjoys comparable computational efficiency with the quickest methods. Thus, with massive non-stationary training data, the ProSpar-GP appears to be capable of leveraging the learned local non-stationary structure for efficient and effective predictive modeling.

% For $d =5$, ProSpar outperforms competitors for both RMSE and CRPS.  Finally, for $D= 10$ ProSpar clearly dominates all other methods by a wide margin. Dtandard inducing point method SVGP underperforms across all sample size and dimension settings. 
% The density of points in low dimensions settings make nearest neighbor conditioning sets very informative regardless of the response function shape.

\section{Application: Surrogate Modeling of Satellite Drag}
\label{sec:app}

We now explore the effectiveness of the ProSpar-GP in a surrogate modeling application on emulating satellite drag; further details in \cite{sun}. Such modeling is needed for avoiding disastrous satellite collisions (see, e.g., one such collision in \citep{achenbach2009debris}) in low earth orbit, which is becoming increasingly crowded. A critical factor to avoiding collisions, as identified by the Committee for the Assessment of the U.S. Air Forces Astrodynamics Standards \citep{national2012continuing}, is the careful modeling of interactions between atmosphere and satellite, specifically with respect to drag coefficients. With recent developments in scientific computing, these drag coefficients can now be accurately simulated via sophisticated physical models, over a broad range of temperature conditions, velocities, satellite geometries and orientations. There is, however, a key limitation: such simulators are too computationally expensive to run for real-time navigation purposes \citep{lawrence2014estimation}.

Surrogate models \citep{gramacy_surrogates_2020} thus provide an attractive solution. The idea is to simulate a carefully-designed set of training data over the parameter space, then use this to train a ``surrogate model'' for emulating the expensive simulator. Prior work \citep{mehta2014modeling,sun} has shown that GPs are quite effective for the surrogate modeling of satellite drag. One requirement for such surrogates is that they should ensure a relative predictive accuracy of 1\% over the parameter space \cite{sun}, to guarantee reliable performance in downstream tasks (e.g., navigation). This becomes more difficult for realistic geometries, which require more parameters and therefore a \textit{larger sample size} from the simulator to achieve the desired accuracy. Furthermore, it has been noted \citep{mehta2014modeling} that the response surface for satellite drag can be highly \textit{non-stationary}. This surrogate modeling application, with massive and non-stationary data, thus provides a nice test problem for the ProSpar-GP.

% Researchers at the Los Alamos National laboratory developed the Test Particle Monte Carlo Simulator to model the drag of various satellite bodies moving through atsmopheric gases \citep{sun}. Their goal was to create a surrogate model that achieves 1\% RMSPE using as few runs of the expensive simulator as possible. 

For our experiments, we consider the satellite drag for the Hubble Space Telescope moving through a pure hydrogen gaseous composition; this is known \citep{sun,scaledvecchia} to be a challenging surrogate modeling problem due to its highly non-concave and non-stationary response surface. We adopt the two million simulation runs from \cite{sun}, which were generated from a Latin hypercube design using the test particle Monte Carlo program developed at Los Alamos National Laboratory. This simulator has a total of $d=8$ parameters, including satellite velocity, yaw and pitch; details can be found in Table 1 of \cite{sun}. We then draw the training and testing data randomly from these simulation runs, with the training data having (large) sample sizes of $n = 1.00 \times 10^5$, $1.75\times 10^5$ and $2.50 \times 10^5$, and the testing set fixed at $1.75 \times 10^5$ samples. The same methods are compared as in earlier numerical experiments, with the nugget terms for SVecGP, VDGP and HLaGP fixed at $10^{-4}$, as recommended in \citep{sun,sauer2022vecchiaapproximated}. As before, all methods are replicated 10 times for each sample size setting, except for the VDGP, which is replicated 3 times due to its high computational cost. The latter is also not applied to the $n = 2.50 \times 10^5$ setting, where a single replication requires several days of runtime.

% We allow SVGP, GPOE, and ProSpar to learn the noise variance terms. We generally have found that SVecGP, VDGP, and HLaGP performance degrades when simultaneously estimating the noise variance parameter. 

\begin{figure}[!t]
    \centering
    \includegraphics[width = \textwidth]{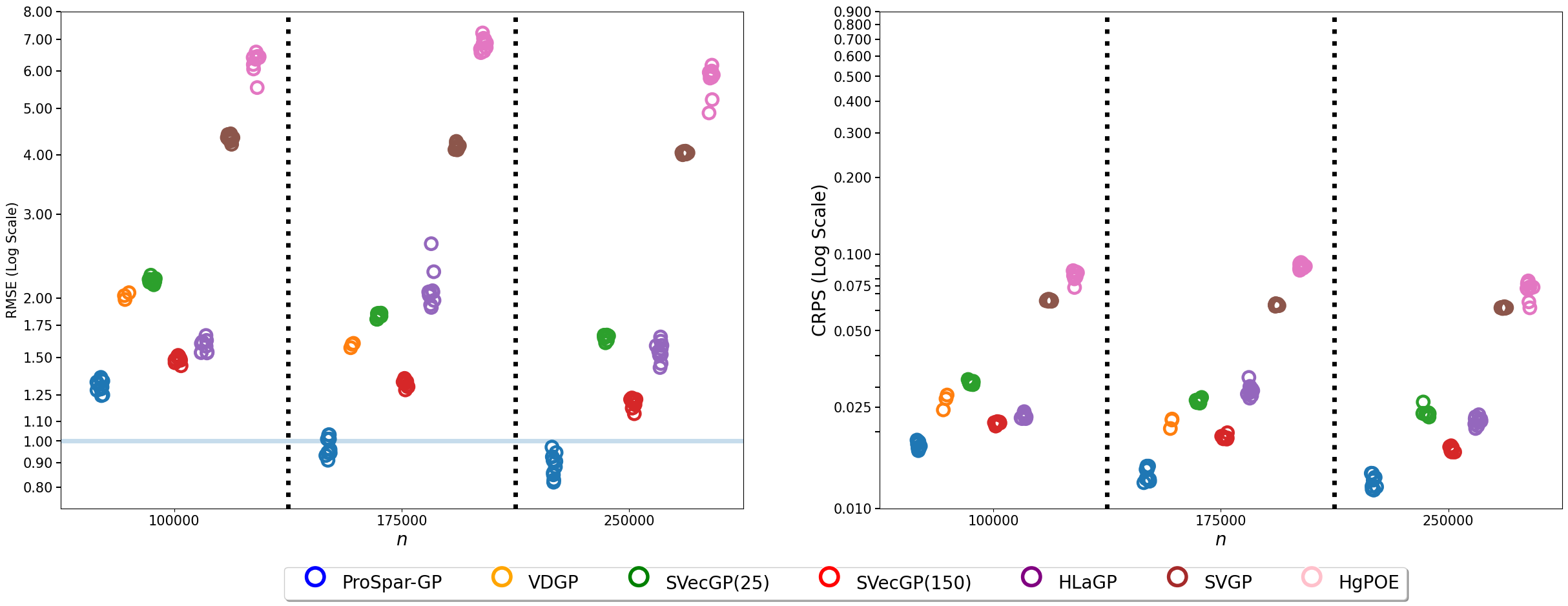}
    \caption{Predictive metrics (left: RMSE; right: CRPS) of the compared methods in the satellite drag surrogate modeling application, for different sample sizes $n$. The desired 1\% RMSE is marked on the left plot by the blue horizontal line.}
    \label{fig:sat_results}
\end{figure}

Figure \ref{fig:sat_results} shows the RMSE and CRPS of the compared methods, with the blue line marking the desired 1\% RMSE threshold for the surrogate model. We see that the SVGP and HgPOE again yield mediocre performance at all sample sizes $n$; this is in line with earlier numerical experiments, where both methods performed poorly in the presence of non-stationarity for moderate to high dimensions. The local-neighbor methods (VDGP, SVecGP, HLaGP) provide improved performance in terms of RMSE and CRPS, but fall short of the desired 1\% RMSE requirement even with $n=2.50 \times 10^5$ training samples. This can be restrictive for the surrogate modeling application, where each simulation run (i.e., data point) from the simulator is computationally expensive; one would thus prefer a surrogate model trained using a smaller training sample size $n$ that satisfies the desired accuracy requirement. The proposed ProSpar-GP offers considerably improved predictions over existing methods, both in terms of RMSE and CRPS. It also addresses the aforementioned limitation: with $n=1.75 \times 10^5$ samples, the ProSpar-GP achieves the 1\% RMSE threshold for the surrogate model, and with $n=2.50 \times 10^5$ samples, this error dips well below the desired threshold. Thus, with careful identification and integration of non-stationary structure within a product-of-experts framework, the ProSpar-GP allows for accurate surrogate modeling of satellite drag with considerably fewer evaluations from the expensive simulator.

Table \ref{tab:surr_time} summarizes the total processing time (in minutes) required for surrogate model training and prediction with different sample sizes $n$. As before, we see that the ProSpar-GP enjoys comparable computing times with the quickest methods, namely, SVecGP(25) and HgPOE. The latter two methods, however, yield considerably worse predictive performance, with RMSE well above the required 1\% threshold (see Figure \ref{fig:sat_results}) for surrogate modeling. Other existing methods require higher training runtimes with considerably higher prediction errors, which is undesirable. Thus, with training and prediction runtimes factored in, the ProSpar-GP appears to considerably outperform its surrogate modeling competitors for this challenging satellite drag application.

% GPOE performed extremely poorly on this emulation task, so we do not display its results as its inclusion makes the graph hard to read. We see that ProSpar outperforms competitors and alone achieves the desired $1\%$ benchmark beginning at $n= 175,000$. 

\begin{table}[!t]
\small
    \centering
\begin{tabular}{cccccccc}
\toprule
\multicolumn{8}{c}{\textbf{Satellite Drag Surrogate Modeling} ($d=8$ parameters)} \vspace{0.1cm}\\
$n$ & ProSpar-GP & VDGP & SVecGP($25$) & SVecGP($150$) & HLaGP & SVGP & HgPOE \\
\midrule
$1.00 \times 10^5$ & 32.18 & 5430.30 & 1.68 & 59.59 & 460.85 & 30.61 & 2.53 \\
$1.75 \times 10^5$ & 35.27 & 5794.89 & 1.21 & 65.23 & 257.97 & 51.87 & 4.39 \\
$2.50 \times 10^5$ & 9.34 & 2089.28 & 1.05 & 33.45 & 46.10 & 14.71 & 2.78 \\
\bottomrule
\normalsize
\end{tabular}
    \caption{Total processing time (in minutes) required for model training and prediction of the compared methods in the satellite drag surrogate modeling application, for different sample sizes $n$. }
\label{tab:surr_time}
\end{table}

\section{Conclusion}
\label{sec:conc}

We introduced in this paper a novel Product of Sparse Gaussian Process (ProSpar-GP) method for scalable GP modeling with massive non-stationary datasets. While there is a notable body of work on scalable methods for GP modeling with big data, we showed that such methods may yield poor approximations when the underlying response surface is non-stationary, particularly in moderate to high dimensions. The ProSpar-GP addresses this critical limitation via a carefully-constructed product-of-expert formulation of sparse GP experts, which leverages an efficient variational inference procedure for optimizing different experts (i.e., its inducing points and length-scale parameters) within local regions of non-stationarity. We further showed that the ProSpar-GP enjoys Kolmogorov consistency, an important ingredient for ensuring stability for variational inference of the procedure. We then demonstrated the improved performance of the ProSpar-GP over the state-of-the-art (particularly in moderate to high dimensions), in a suite of numerical experiments and a surrogate modeling application for simulating satellite drag.

Given the promising results in this paper, there are many fruitful avenues for impactful future work. For high-dimensional systems, particularly in the physical sciences, one often expects the presence of low-dimensional structure, e.g., in the form of manifold embeddings \citep{zhang2022gaussian,li2023additive} or structured sparsity \citep{tang2023hierarchical}. One direction is thus the integration of such low-dimensional structure within each local expert in the ProSpar-GP, which can better refine non-stationary predictions in high dimensions. Another direction is the application of the ProSpar-GP for surrogate modeling in modern scientific problems with massive datasets, e.g., the real-time control of engines in unmanned aerial vehicles \cite{narayanan2023physics}.

\if1\blind{
\noindent \textbf{Acknowledgements}: The authors gratefully acknowledge funding from NSF CSSI 2004571, NSF DMS 2210729, NSF DMS 2316012 and DE-SC0024477. We also thank the JETSCAPE collaboration (\url{https://jetscape.org/}) for insightful conversations and discussions.
}
\fi

% This paper provides a method for non-stationary GP regression by aggregating multiple sparse GP experts into a global stochastic model via  product of experts. Aggregation of these experts into a coherent global probabilistic model enables scalable training and hyper-parameter selection via stochastic variational inference. We demonstrated superior performance of our method in non-stationary regression tasks in moderate to higher dimensions. 

\FloatBarrier
\spacingset{1.2}
\fontsize{11pt}{11pt}\selectfont
\bibliography{bib}

\pagebreak
\normalsize

\pagebreak

\section*{Supplementary Material}

% \subsection{Analytical Posterior of ProSpar}
% We can calculate the analytical marginal likelihood and posterior of ProSpar GP. 

% \subsection{Satellite Run times}
% \label{sup:sattelite}

% \begin{table}[]
% \begin{tabular}{lrrrrrrr}
% \toprule
% Method & ProSpar-GP & VDGP & SVecGP(25) & SVecGP(150) & HLaGP & SVGP & HgPOE \\
% n &  &  &  &  &  &  &  \\
% \midrule
% 100000 & 32.18 & 5430.30 & 1.68 & 59.59 & 460.85 & 30.61 & 2.53 \\
% 175000 & 35.27 & 5794.89 & 1.21 & 65.23 & 257.97 & 51.87 & 4.39 \\
% 250000 & 54.05 &  & 1.37 & 62.12 & 260.58 & 73.94 & 7.00 \\
% \bottomrule
% \end{tabular}
% \caption{Total processing time in minutes for training and prediction for Sattelite drag emulation problem. VDGP not included for $n = 250000$ for computational reasons.}
% \end{table}

% \begin{table}[]
%     \centering
%     \begin{tabular}{lrrrrrrr}
%     \toprule
%     d & ProSpar & VDGP & SVecGP $25$ & SVecGP $150$ & HLaGP & SVGP & GPOE \\    \midrule
%     2 & 5.90 & 432.58 & 0.46 & 17.39 & 52.52 & 14.81 & 3.57 \\
%     5 & 6.67 & 1069.63 & 0.62 & 37.61 & 76.46 & 14.84 & 3.08 \\
%     10 & 8.47 & 2089.28 & 2.00 & 64.79 & 45.42 & 14.71 & 2.78 \\
%     \bottomrule
%     \end{tabular}
%     \caption{Total runtimes in minutes for G-function emulation task for $n = 50,000$}
%     \label{tab:my_label}
% \end{table}

\subsection*{Proof of Proposition \ref{prop:cons}}
\label{sup:proof}
For Kolmogorov consistency to hold, we would need to show both conditions in Proposition \ref{prop:cons} hold. To see that the first condition is true, note that:
\begin{align*}
 &p(\bm{f}) = p(\bm{f}(\bm{x}_1), \dots, \bm{f}(\bm{x}_n)) \sim \w{N}(\bm{f}; \bm{0}, \bs{\Sigma}), \\
 &  \bs{\Sigma} = \bs{\Lambda} + \sum_{j=1}^J \bs{\Lambda} \bm{A}_j \bs{\Lambda}_j^{-1} \bm{K}_{n,m_j}^{[j]} \bm{R}_j^{-1} \bm{K}_{m_j,n}^{[j]}\bs{\Lambda}_j^{-1} \bm{A}_j  \bs{\Lambda}.
\end{align*}
Note that any permutation to the ordering of observations $\{(\bm{x}_{\pi(i)}, f(\bm{x}_{\pi(i)}))\}_{i=1}^n$ simply permutes the rows/columns of the matrices $\bs{\Lambda}$, $\bs{\Lambda}_j$, $\bm{A}_j$ and $\bm{K}_{n,m_j}^{[j]}$. Because $\bs{\Lambda}$, $\bs{\Lambda}_j$, $\bm{A}_j$ are diagonal and $\bm{R}_j^{-1}$ does not depend on the observations, such a reordering results in a commensurate reordering of the elements in  $\bs{\Sigma}$. Thus, the density remains unchanged, as desired.

The second condition follows from the marginalization property of the multivariate normal distribution. Here, $\bs{\Lambda}$ is diagonal, and given the expert parameters, calculation of the $i$-th diagonal element only requires the $i$-th observation. In addition, the covariance matrix of any $n_c$-subset of the $n$ observations only requires entries from the $n_c$ corresponding rows/columns of the matrices $\bs{\Lambda}$, $\bs{\Lambda}_j$, $\bm{A}_j$ and $\bm{K}_{n,m_j}^{[j]}$. Thus, inclusion or exclusion of an observation does not affect the parameters of any other latent function values, and we can simply marginalize out $\bm{f}_2$ as with any standard multivariate normal density.  

\end{document}